\documentclass[journal]{IEEEtran}
\usepackage{subfigure}
\usepackage{cite}
\usepackage{hyperref}
\usepackage{csquotes}
\usepackage{balance}
\usepackage{color}
\usepackage{subfigure}
\usepackage{graphicx, hyperref, url}
\usepackage{tabularx}
\usepackage{tabulary, booktabs}
\usepackage{longtable}
\usepackage{lipsum}
\usepackage[dvipsnames]{xcolor}
\usepackage{amsmath}
\usepackage{array}
\usepackage[T1]{fontenc}
\usepackage[official]{eurosym}
\newcommand{\etal}{\textit{et al. }}

\begin{document}

\title{Quantum Internet- Applications, Functionalities, Enabling Technologies, Challenges, and Research Directions}

\author{Amoldeep Singh, Kapal Dev, \IEEEmembership{Member, IEEE}, Harun Siljak, \IEEEmembership{Senior Member, IEEE}, Hem Dutt Joshi, \IEEEmembership{Member, IEEE}, Maurizio Magarini, \IEEEmembership{Member, IEEE}.


\thanks{Amoldeep Singh is with the department of Electronics and Communication Engineering, Thapar Institute of Engineering and Technology, Patiala, Punjab, India, (e-mail: abharee\textunderscore phd17@thapar.edu).}

\thanks{Kapal Dev is with Department of institute of intelligent systems, University of Johannesburg, South Africa, (e-mail: kapal.dev@ieee.org).}

\thanks{Harun Siljak is with the Department of Electronic and Electrical Engineering, Trinity College Dublin, Ireland, (e-mail: harun.siljak@tcd.ie).}

\thanks{Hem Dutt Joshi is with the department of Electronics and Communication Engineering, Thapar Institute of Engineering and Technology, Patiala, Punjab, India, (e-mail: hemdutt.joshi@thapar.edu).}
\thanks{Maurizio Magarini is with the Dipartimento di Elettronica, Informazione e Bioingegneria, Politecnico di Milano, 20133 Milan, Italy, (e-mail: maurizio.magarini@polimi.it).}
}

\maketitle

\begin{abstract}
The advanced notebooks, mobile phones, and internet applications in today's world that we use are all entrenched in classical communication bits of zeros and ones. Classical internet has laid its foundation originating from the amalgamation of mathematics and Claude Shannon's theory of information. But today's internet technology is a playground for eavesdroppers. This poses a serious challenge to various applications that relies on classical internet technology. This has motivated the researchers to switch to new technologies that are fundamentally more secure. Exploring the quantum effects, researchers paved the way into quantum networks that provide security, privacy, and range of capabilities such as quantum computation, communication, and metrology. The realization of Quantum Internet (QI) requires quantum communication between various remote nodes through quantum channels guarded by quantum cryptographic protocols. Such networks rely upon quantum bits (qubits) that can simultaneously take the value of zeros and ones. Due to the extraordinary properties of qubits such as superposition, entanglement, and teleportation, it gives an edge to quantum networks over traditional networks in many ways. But at the same time transmitting qubits over long distances is a formidable task and extensive research is going on satellite-based quantum communication, which will become a breakthrough in physically realizing QI in near future. In this paper, QI functionalities, technologies, applications and open challenges have been extensively surveyed to help readers gain a basic understanding of the infrastructure required for the development of the global QI.    
\end{abstract}


\IEEEpeerreviewmaketitle

\section{OUTLINE}
\label{sec: Sec. I}
\IEEEPARstart{Q}{uantum Internet} \textcolor{black}{(QI)} is trending internet technology that facilitates quantum communication through quantum bits (qubits) among remote quantum devices or nodes. Such a technology will work in synergy with classical internet to overcome the limitations posed by traditional interconnect technologies, as was highlighted in~\cite{hanzo2012wireless}. 
This novel technology is administered by the laws of quantum mechanics, where one of the laws states that it is impossible to measure a property of a system without changing its state. Consequently, qubits cannot be copied and any attempt to do so will be detected, thus making the communication more secure and private \cite{wehner2018quantum}. Due to the extraordinary properties of qubits, it gives an edge to~\textcolor{black}{QI} over traditional internet in many ways~\cite{van2012quantum}.

Qubits also show weird phenomena of quantum \textit{entanglement} \cite{nielsen2002quantum}, where qubits at remote nodes are correlated with each other. This correlation is stronger than ever possible in the classical domain. Entanglement is inherently private, as it is not possible, because of no-cloning \cite{wootters1982single}, for a third qubit to be entangled with either of the two entangled qubits. Therefore, with this weird quantum effect, \textcolor{black}{QI} could open up a different galaxy of applications with world-changing potential \cite{komar2014quantum}, \cite{dai2020towards}, \cite{bahder2004quantum}, \cite{ekert1991quantum}, \cite{crepeau2002secure}, \cite{fitzsimons2017unconditionally}, \cite{giovannetti2004quantum}, \cite{gottesman2012longer}. 
As qubits are prone to environmental losses, transmitting qubits over long distances is a challenging task due to \textit{decoherence} \cite{brandt1999qubit}. Therefore, extensive research is going on to acheive long-haul quantum communication \cite{valivarthi2016quantum}, \cite{ren2017ground}. To mitigate this noise, various techniques are employed such as Quantum Error-Correcting (QEC) codes~\cite{knill1997theory} and fault-tolerant techniques~\cite{fowler2010surface}, \cite{munro2012quantum}, \cite{muralidharan2014ultrafast}. Consequently, quantum communication can bear a finite amount of noise, unlike classical communication.

\begin{figure}[!t]
\centering
\includegraphics[width=0.5\textwidth]{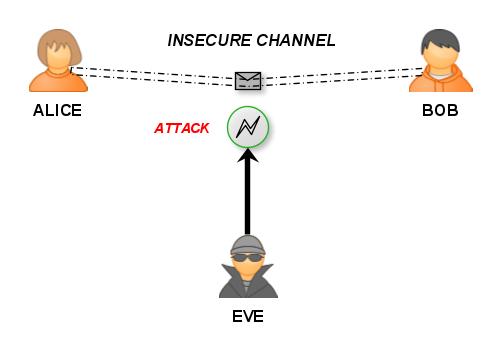}
\caption{Communication between Alice and Bob intercepted by \textit{Eve}. Here channel is insecure which means that information is not encrypted by a \textit{cipher}, thus vulnerable to attacks by \textit{Eve}.}
    \label{fig:Fig 1}
\end{figure}

In modern era, where internet plays a vital role in everyday life, secure communication is the main concern ensuring privacy between two communicating parties without any eavesdropping. These issues are readily addressed by cryptographic techniques wherein data is protected with a private key. \textcolor{black}{New potential in cryptography emerges with advances in \textit{quantum cryptography} that exploits quantum mechanical principle of \textit{no-cloning}. These features play a vital role in providing secrecy and integrity to the data to be communicated, thus making messages unintelligible to any unauthorized party.} If a malevolent third party, such as \textit{Eve}, as shown in \textbf{Fig. \ref{fig:Fig 1}}, eavesdrops on this key distribution, privacy of the communication will be compromised. This problem is addressed by the best-known application of quantum networking, \textcolor{black}{i.e. Quantum Key Distribution (QKD)}, which provides secure access to computers on the cloud \cite{bennett1992quantum} utilizing various protocols. QKD networks are commercially available \cite{diamanti2016practical}, \cite{inagaki2013entanglement}, and are studied and deployed covering metropolitan distances \cite{peev2009secoqc}, \cite{sasaki2011field}, \cite{stucki2011long}, \cite{wang2014field}. Long distance QKD networks with trusted nodes are also currently possible \cite{peev2009secoqc}, \cite{courtland2016china}. Therefore, ultra-secure QKD protocols are developed based on well-accepted laws that govern quantum physics for sharing secret keys among two parties \cite{ekert1991quantum}.  So, a lot of interdisciplinary effort is required by research groups working on this novel technology to make it a reality \cite{caleffi2020rise}, \cite{pirandola2016physics}.



\subsection{Quantum race}
\label{sec: Sec. I-A}
A tremendous amount of effort is put together by scientists to realize global \textcolor{black}{QI}. As a result, infrastructures to realize this technology are extensively studied and proposed. For instance, \textcolor{black}{manufacturing of quantum computers is speeding up} as various tech giants such as IBM, Google, Intel, and Alibaba have already started working on it. A 50-qubits processor has been built and tested by IBM in November 2017 \cite{castelvecchi2017ibm}. Google has announced a 72-qubit processor \cite{kelly2019operating} and recently, towards the end of 2019, the company has claimed to enter a new era of incredible computational power by achieving something called ‘quantum supremacy’, i.e. solving a computational problem that has not had a classical computing solution obtainable in a reasonable time span \cite{preskill2012quantum}. Intel and Alibaba are actively working on technologies to develop double-digit-qubits processors. In this quantum computing race, future quantum computers will come into force having disruptive potential to tackle complex problems such as understanding photosynthesis, improvising catalysts for formulating renewable fuels~\cite{cao2019quantum}, \cite{bourzac20174}, and complex Artiﬁcial Intelligence (AI) systems \cite{gil20201}.

In the interim, European countries are also actively involved in building \textcolor{black}{QI} and, to boost the research, the European Commission has hurled a ten-year \euro{1-billion} leading project~\cite{gibney2016billion}. In 2018, the European Union has announced its first grant of \euro{132-million} as an initiative in the quantum flagship for the following three years to speed up the quantum research \cite{cartlidge2018europe}. Stephanie Wehner group at the Delft University of Technology in the Netherlands has a strong vision for \textcolor{black}{QI} technology. Their team at Delft is working to assemble the first authentic quantum network, which is planned to link four urban areas in the Netherlands \cite{QIA}. They are coordinating in a large venture called Quantum Internet Alliance (QIA) and together with other scientists are trying to reach their goal to develop a blueprint for a pan-European entanglement-based \textcolor{black}{QI} \cite{QIA}. This blueprint will set the stage for a strong European \textcolor{black}{QI} industry. Outside Europe, the US government has earmarked nearly 1.3 billion US dollars in funding for quantum research via its National Quantum Initiative Act, which has to run from 2019 to 2023~\cite{monroe2019us}. In the year 2017, a \textcolor{black}{research group from China} has successfully demonstrated and observed a satellite-based distribution of entangled photon pairs to two separate ground stations on the earth, and with the successful launch of the Micius satellite, they have already invested in the first step to realize \textcolor{black}{QI} as a global network \cite{yin2017satellite}.

\begin{figure}[!t]
\centering
\includegraphics[width=0.48\textwidth]{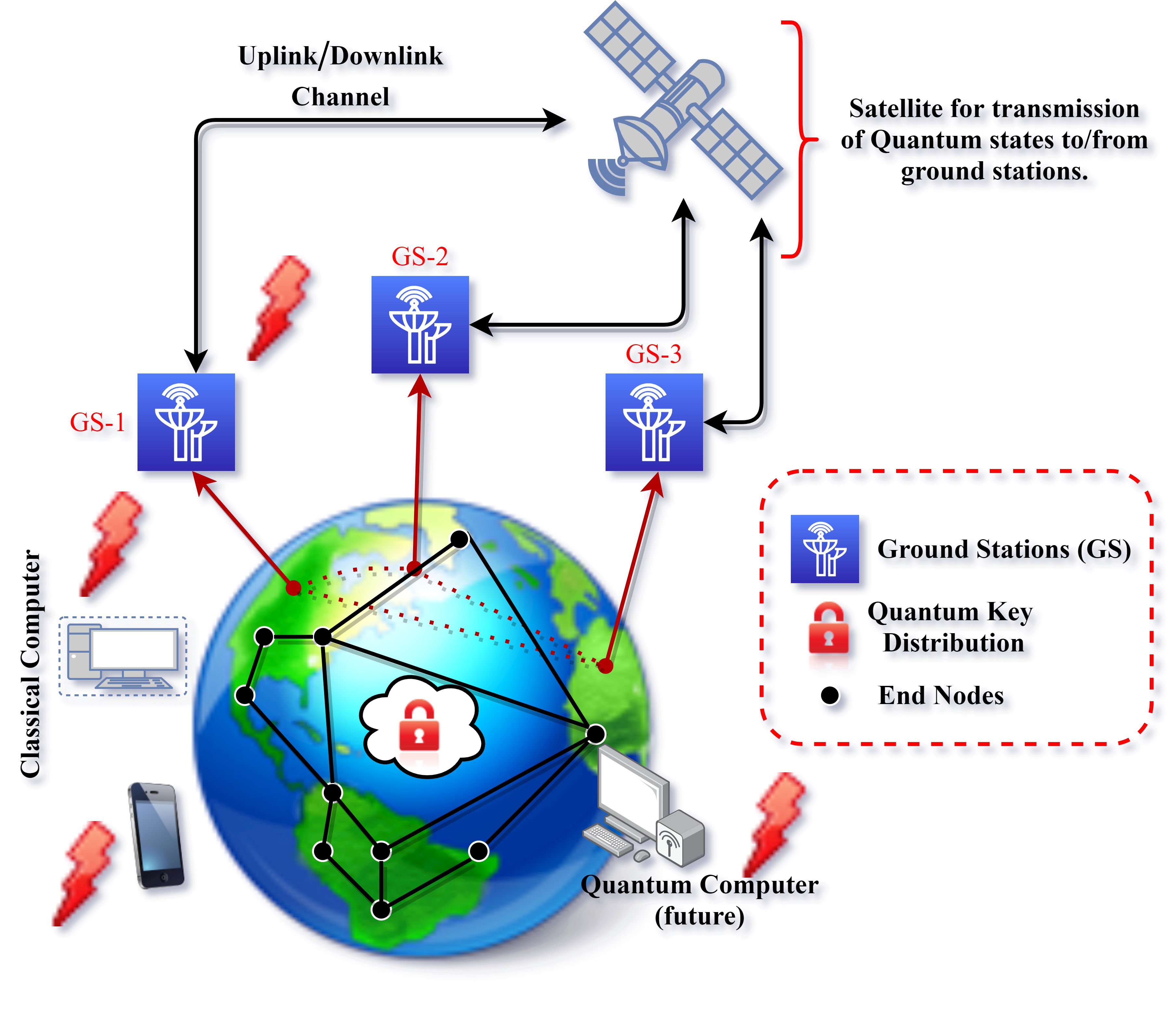}
\caption{Vision of future quantum internet working in synergy with classical internet.}
\label{fig:Fig 2}
\end{figure}

 \
 
\begin{figure*}[ht!]
\centering
  \includegraphics[width=5in, height=14.8cm]{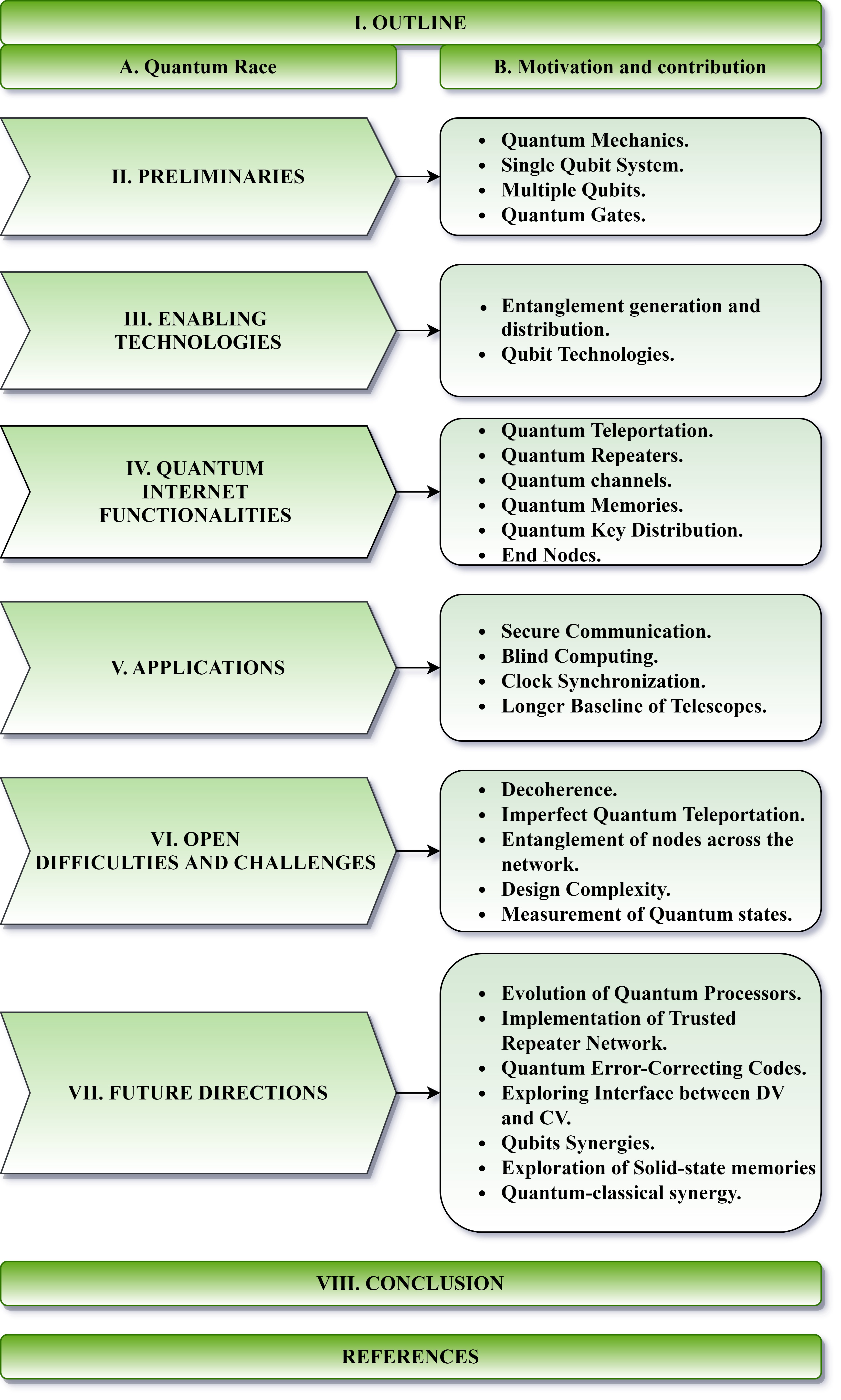}
\caption{Structure of the paper}
  \label{fig:Fig 3}
\end{figure*}


\textcolor{black}{It is envisioned that a future QI will somewhat look like as seen in \textbf{Fig. \ref{fig:Fig 2}}. It will have a network of remote nodes interconnected with each other with the help of multiparty entanglement and quantum teleportation using fiber-based or free space channels. The quantum effects of quantum mechanics will make this network secure using various QKD protocols. QEC codes will be required to encode the qubits carrying the information that will protect them from decoherence and environmental effects. With advancements in multi qubit processors, full scale quantum computers will come into force that, along with existing classical infrastructure, will revolutionize the world with inherently secure global QI.}

\subsection{Motivation and Contribution}
\label{sec: Sec. I-B}
There are some specific surveys and review papers available that include the research work accomplished in this technology, but lacks a generalized review. A brief comparison of this survey with other published surveys on quantum networking is presented in \textbf{Table \ref{tab: I}}. \textcolor{black}{It is worth noting that several published surveys are specific to certain concepts such as quantum computing \cite{gyongyosi2019survey}, quantum cryptography \cite{pirandola2006quantum}, quantum teleportation \cite{pirandola2015advances}, \cite{cacciapuoti2020entanglement}, quantum channel capacities \cite{gyongyosi2018survey}, quantum key distribution \cite{xu2020secure}, \cite{scarani2009security}, quantum entanglement \cite{adesso2007entanglement}, and satellite based quantum communication \cite{bedington2017progress}, \cite{hosseinidehaj2018satellite}. Motivated by this, we have adopted a generalized approach targeting the wide audience to appreciate the advantages of quantum communication over the classical one, with the following contributions: 
\begin{itemize}
    \item The basic concepts of quantum mechanics that are required to have an in-depth knowledge of quantum communication are put forth.
    \item We provide enough tutorial content by highlighting the functionalities of QI, followed by possible applications it can offer, and by underlining the challenges it can face with.
    \item We cover all the components of QI explicitly so that even the readers without quantum related background can easily understand the preliminaries governing this novel technology.
    \item We bring all the components of QI together, which are required to build the global quantum networks, by outlining their historical developments with communication perspective.
\end{itemize}}
   
\newcolumntype{Q}[1]{>{\centering\arraybackslash}p{#1}}
\begin{table*}[ht!]
\caption{Comparison of this survey with other related surveys}
   \centering
\begin{tabular}{|Q{2.5cm}||Q{1cm}|Q{1cm}|Q{1cm}|Q{1cm}|Q{1cm}|Q{1cm}|Q{1cm}|Q{1cm}|Q{1cm}|Q{1.5cm}|}
 \hline
 Topics& Basics of Quantum Mechanics & Qubit Technologies &Quantum entanglement& Quantum Teleportation & Quantum Repeaters &Quantum Channels & Quantum Memories & QKD &End Nodes &Quantum error correction \\
 \hline
 \hline
 Laszlo Gyongyosi and Sandor Imre (2019) \cite{gyongyosi2019survey}   & $\surd$&    &  &  & & &$\surd$& &&$\surd$ \\
 \hline
 Stephanie Wehner (2018) \etal \cite{wehner2018quantum}   & &&$\surd$ &$\surd$  &$\surd$ &$\surd$ &$\surd$&$\surd$ &$\surd$&$\surd$ \\
 \hline
 Robert Bedington (2017) \etal \cite{bedington2017progress}   &  && $\surd$& $\surd$&$\surd$&&&$\surd$&&$\surd$\\
 \hline
 Valerio Scarani \etal (2009) \cite{scarani2009security}   &  & & $\surd$& $\surd$&$\surd$&$\surd$&&$\surd$&&$\surd$\\
 \hline
 Angela Sara Cacciapuoti \etal (2020) \cite{cacciapuoti2020entanglement} & $\surd$ & & $\surd$& $\surd$&$\surd$&& $\surd$&$\surd$&&$\surd$\\
 \hline
 Stefano Pirandola \etal (2015) \cite{pirandola2015advances} & &$\surd$ & $\surd$& $\surd$&$\surd$&&&$\surd$&&\\
 \hline
 Gerardo Adesso and Fabrizio Illuminati (2007) \cite{adesso2007entanglement} & $\surd$ && $\surd$& $\surd$&$\surd$&&&$\surd$&&\\
 \hline
 Hoi-Kwong Lo \etal (2014) \cite{lo2014secure} &  & & $\surd$& $\surd$&&&&$\surd$&&$\surd$\\
 \hline
 Laszlo Gyongyosi \etal (2018) \cite{gyongyosi2018survey}  & $\surd$ & & $\surd$& &&$\surd$&&$\surd$&&$\surd$ \\
 \hline
Feihu Xu \etal (2020) \cite{xu2020secure} & & &$\surd$ & &$\surd$&$\surd$&$\surd$&$\surd$&$\surd$&$\surd$ \\
 \hline
 Stefano Pirandola \etal (2020) \cite{pirandola2020advances} & & &$\surd$ &  &&$\surd$&&$\surd$&$\surd$ &$\surd$\\
 \hline
 Zunaira Babar \etal (2018) \cite{babar2018duality} &$\surd$ & &$\surd$ &  & &&&&&$\surd$\\
 \hline
 Nedasadat Hosseinidehaj \etal (2019) \cite{hosseinidehaj2018satellite} &$\surd$ & &$\surd$ &  & &$\surd$&$\surd$&$\surd$&$\surd$&$\surd$\\
 \hline
 This Survey & $\surd$ &$\surd$& $\surd$& $\surd$&$\surd$&$\surd$&$\surd$&$\surd$&$\surd$&$\surd$
 \\
 
 \hline
\end{tabular}
\label{tab: I}
\end{table*}

\textcolor{black}{The rest of the paper structure is detailed as shown in~\textbf{Fig. \ref{fig:Fig 3}}. Section II starts with the preliminaries of quantum mechanics that are required to understand the basic concepts needed for quantum computation and communication. In Section III, different technologies adopted to represent qubits are listed. This is followed by reviewing the functionalities of \textcolor{black}{QI} in Section IV, which includes quantum teleportation, quantum repeaters, quantum channels, quantum memories, QKD, and end nodes. In Section V, some of the applications that have been discovered theoretically by researchers are listed and reviewed. After that, in Section VI, challenges faced in realizing \textcolor{black}{QI} are analysed such as decoherence, imperfections faced by quantum teleportation process, entanglement generation, and distribution of quantum states. Finally, in Section VII, future perspectives are discussed followed by concluding remarks and research directions for the realization of the global \textcolor{black}{QI}.}

\section{PRELIMINARIES}
\label{sec: Sec. II}
In this section we will study the basics of quantum mechanics with emphasis on qubits that carry the quantum information between the remote nodes. We have discussed the mathematics behind single qubit and multi qubit systems, state vector representations on Bloch sphere, followed by quantum gates for local operations and manipulation of quantum circuits.

\subsection{Quantum Mechanics}
\label{sec: Sec. II-A}
Quantum mechanics has been an irreplaceable piece of science lately and is applied with colossal success in many fields including physics, chemistry, AI, life sciences, and military applications. It is termed as a mathematical framework, or set of rules, for the development of physical hypotheses by exploiting the properties of single quantum systems involving qubits. For harnessing the power of quantum mechanics in various applications, the study of the qubit and of the single quantum system is of utter importance. It has the potential to offer secure communications \cite{hu2016experimental} that can make them cryptographically as secure as the One-Time-Pad (OTP) \cite{vernam1926cipher}, unlike today’s classical communication that is vulnerable to hacker attacks. For better understanding the quantum effects of quantum mechanics, certain theorems \cite{kanamori2006short} were postulated by physicists to get a clear view of properties of qubits.   
\subsubsection{State space and superposition principle}
\label{sec: Sec. II-A-1}
\begin{figure}[!t]
\centering
\includegraphics[width=0.5\textwidth]{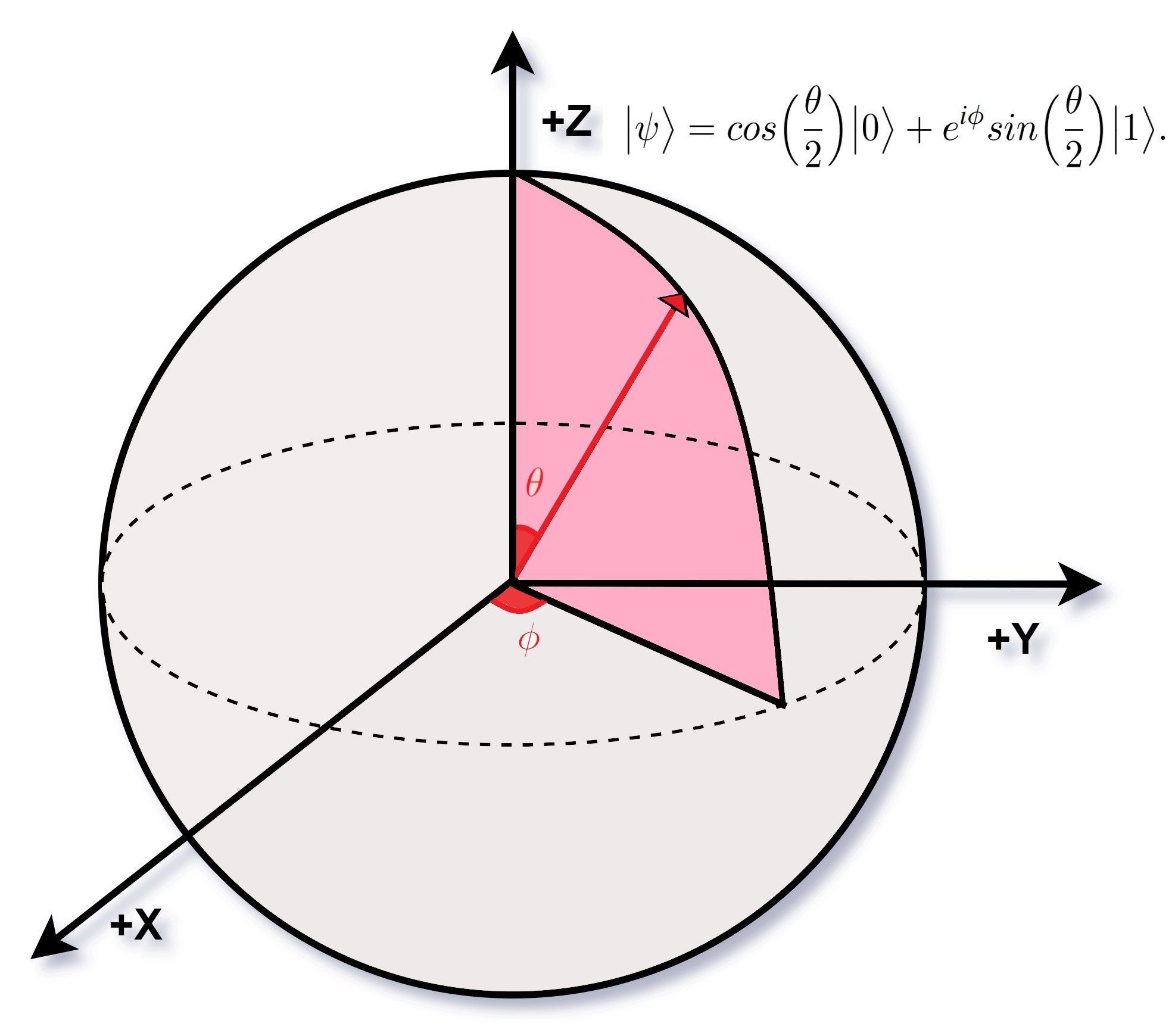}
\caption{Bloch sphere: Geometrical representation of pure quantum state of a qubit. Here any state $\big|\psi\big\rangle$ = $\alpha\big|0\big \rangle$ + $\beta\big|1\big\rangle$ can be shown by a point on the surface of the sphere, where $\alpha$ = cos$\big(\frac{\theta}{2}\big)$ and $\beta$ = $e^{\textit{i}\phi}$sin$\big(\frac{\theta}{2}\big)$.}
    \label{fig:Fig 4}
\end{figure}

The system of qubits is associated with complex Hilbert space, which is an isolated or closed quantum physical system, also known as the \textit{state space} of the system. A unit vector completely describes the state vector of a system in its state space. A qubit is the simplest quantum mechanical system that can be represented geometrically by a Bloch sphere, as depicted in \textbf{Fig. \ref{fig:Fig 4}}. Any \textcolor{black}{\textit{pure} quantum state} $\big|\psi\big\rangle$ can be represented by a point on the surface of the Bloch sphere with spherical coordinates of \textcolor{black}{$\theta$ as polar angle, i.e. angle that a line makes with Z-axis, and $\phi$ as an azimuthal angle, i.e. angle that line makes with X-axis \cite{pati2000minimum}:} 
\begin{equation}\label{eq1}
 \big|\psi\big\rangle=\cos\Big(\frac{\theta}{2}\Big)\big|0\big \rangle+e^{\textit{i}\phi}\sin\Big(\frac{\theta}{2}\Big)\big|1\big\rangle,
\end{equation}
where “$\big|\cdot\big\rangle$” is called a “ket-vector” in \textit{Dirac notation}, which physically represents the \textcolor{black}{pure} quantum state of the qubit. The angle \textcolor{black}{$\phi$ represents the \textit{phase} of a quantum state $\big|\psi\big\rangle$.} Every individual state in the Bloch sphere is represented by a two-dimensional (2D) vector, where $\big|0\big\rangle$ and $\big|1\big\rangle$ are its basis vectors that are orthogonal to each other.
\textcolor{black}{States that can be represented by a “ket-vector” are known as a \textit{pure quantum states}. In comparison, \textit{mixed states} are those states that are not true quantum states, resulting in probabilistic results when measured with all basis. It is a system with weak state or whose state is not fully defined, written in infinitely many different ways as probable outcomes of well-defined pure states \cite{cacciapuoti2020entanglement}, \cite{ballentine2009states}. The pure state, given in equation (\ref{eq1}), upon measurement results in either state $\big|0\big\rangle$ or $\big|1\big\rangle$, as depicted by \textit{Copenhagen interpretation} \cite{wimmel1992quantum} .} 

\textcolor{black}{While in classical communication, the transmission of information takes place in the form of bits taking either ‘0’ or ‘1’ value, the quantum communication on the other hand uses qubits that can be in a \textit{superposition} of the two basis states simultaneously taking the value of ‘0’ and ‘1’ \cite{cacciapuoti2019quantum}. The result of measurement of such states is not definite $\big|0\big\rangle$ or definite $\big|1\big\rangle$. The principle of superposition can be best understood by visualizing the polarization state of a photon. At 45 degree of polarization, a photon is simultaneously vertically and horizontally polarized, representing both the  states at the same time \cite{van2012quantum}. A famous paradox of superposition principle is \textit{Shrödinger's cat}, where cat is simultaneously dead and alive as a result of random event that may or may not occur \cite{schrodinger1935gegenwartige}.}   


\subsubsection{Measurement of Quantum state}
\label{sec: Sec. II-A-2}
As discussed above, a qubit upon measurement will collapse it into one of the basis states \cite{nielsen2002quantum}. This property of qubits is very well utilized in keeping the communications secure as any measurement of quantum state by \textcolor{black}{\textit{Eve}} will affect its state irreversibly. In (\ref{eq1}), the coefficients of $\big|0\big\rangle$ and $\big|1\big\rangle$ represent the probability amplitudes and their squares represent the probabilities of getting ‘0' and ‘1' upon measurement. 
The value of probability amplitudes of the quantum states greatly affects the outcome of measurement and these can be manipulated by quantum gates that will be discussed in \textbf{Sec. \ref{sec: Sec. II-D}}. Measurements can also be understood with the help of the Bloch sphere, as earlier discussed. If the measurement is performed along the Z-axis, the result will be ‘0' or ‘1'. If it is performed along 
the Y-axis, the result will be $\frac{1}{\sqrt{2}}\big|0\big \rangle+\frac{1}{\sqrt{2}}\textit{i}\big|1\big \rangle$ or $\frac{1}{\sqrt{2}}\big|0\big \rangle-\frac{1}{\sqrt{2}}\textit{i}\big|1\big \rangle$. The positive X-axis is pointing towards the reader, the state is $\frac{1}{\sqrt{2}}\big|0\big \rangle+\frac{1}{\sqrt{2}}\big|1\big \rangle$ and $\frac{1}{\sqrt{2}}\big|0\big \rangle-\frac{1}{\sqrt{2}}\big|1\big \rangle$ and any measurement performed in this direction will result in any one of these states. 

\subsubsection{Quantum No-Cloning}
\label{sec: Sec. II-A-3}
This theorem, which was first \textcolor{black}{formulated} by Wooters, Zurek and Deiks in 1982 \cite{wootters1982single}, states that the quantum state of any particle carrying a qubit in a quantum channel cannot be either copied, amplified or cloned, thus making it a reliable and secure form of communication. Any attempt to clone it will be detected, further activating the QKD protocols \cite{bennett1984quantum}. However, at the same time, no-cloning prevents qubits to be sent over long distances by amplification or sending copies of qubits, as it is done in classical communication, for efficient detection of bits without any loss of information. For this purpose, \textit{quantum repeaters} are studied to solve the problem of long distance communication of qubits, which is the core functionality required for full-scale \textcolor{black}{QI}. 

\subsection{Single Qubit system}
\label{sec: Sec. II-B}
\textcolor{black} {In classical digital communication systems, the information is encoded in the form of binary bits. The bit represents a logical state with one of two possible values, i.e. "1" or "0".}
On the other hand, in quantum communication, instead of bits, qubits are employed to carry quantum information from source to destination. To get a solid vibe for how a qubit can be acknowledged, it might be imagined as a \textit{two-way} system; direction of polarization of a photon, the up or down spin of an electron, or two energy levels of an electron orbiting an atom. For instance, the electron can exist in either ‘ground' or ‘excited' state in an atom, represented by a state $\big|\psi\big\rangle = \alpha\big|0\big \rangle + \beta\big|1\big\rangle$, where $\big|0\big \rangle$ and $\big|1\big\rangle$ corresponds to ground state and excited state, respectively. By sparkling some light on the atom with proper time and energy, it is conceivable to move the electron from the $\big|0\big\rangle$ state to the $\big|1\big \rangle$ state and the other way around. Strikingly, by diminishing the time we sparkle the light, an electron, at first, in the state $\big|0\big \rangle$ can now be moved somewhere between $\big|0\big \rangle$ and $\big|1\big \rangle$, into $\big|\psi\big\rangle$ = $ \frac{1}{\sqrt{2}} \big|0\big \rangle$ + $ \frac{1}{\sqrt{2}}\big|1\big \rangle$ state,  i.e. superposition of basis states. The variables $\alpha$ and $\beta$ are complex numbers representing probability amplitudes, which means that $|\alpha|^{2}$ is the probability of getting $\big|\psi\big\rangle$ = 0 as a result of the measurement on qubit $\big|\psi\big\rangle$ and $|\beta|^{2}$ is the probability of getting $\big|\psi\big\rangle$ = 1 as a result of the measurement on qubit $\big|\psi\big\rangle$.
\textcolor{black}{It must be satisfied that,}
\begin{equation}\label{eq2}
|\alpha|^{2}+|\beta|^{2}=1.
\end{equation}

A classical bit is like a coin resulting in definite outputs of heads or tails. 
\textcolor{black}{Supposing, a qubit with $\alpha$ = $\beta$ = $\frac{1}{\sqrt{2}}$, will be in the state:
\begin{equation}\label{eq3}
\big|\pm\big\rangle=\frac{1}{\sqrt{2}}\big|0\big \rangle\pm\frac{1}{\sqrt{2}}\big|1\big \rangle,
\end{equation}
which gives a result 0 with 0.5 probability and result 1 with 0.5 probability, when measured over the time. This state often plays a vital role in quantum communication and is denoted by $\big|+\big\rangle$ and $\big|-\big\rangle$.} Despite qubit theoretically having the capacity of carrying an infinite amount of information, it has a major flaw that all the quantum states collapse to a single state of 0 and 1 upon measurement \textcolor{black}{and the reason for this characteristic is still unknown.} 
\begin{figure}[!t]
\centering
\includegraphics[width=0.48\textwidth, height=5.5cm]{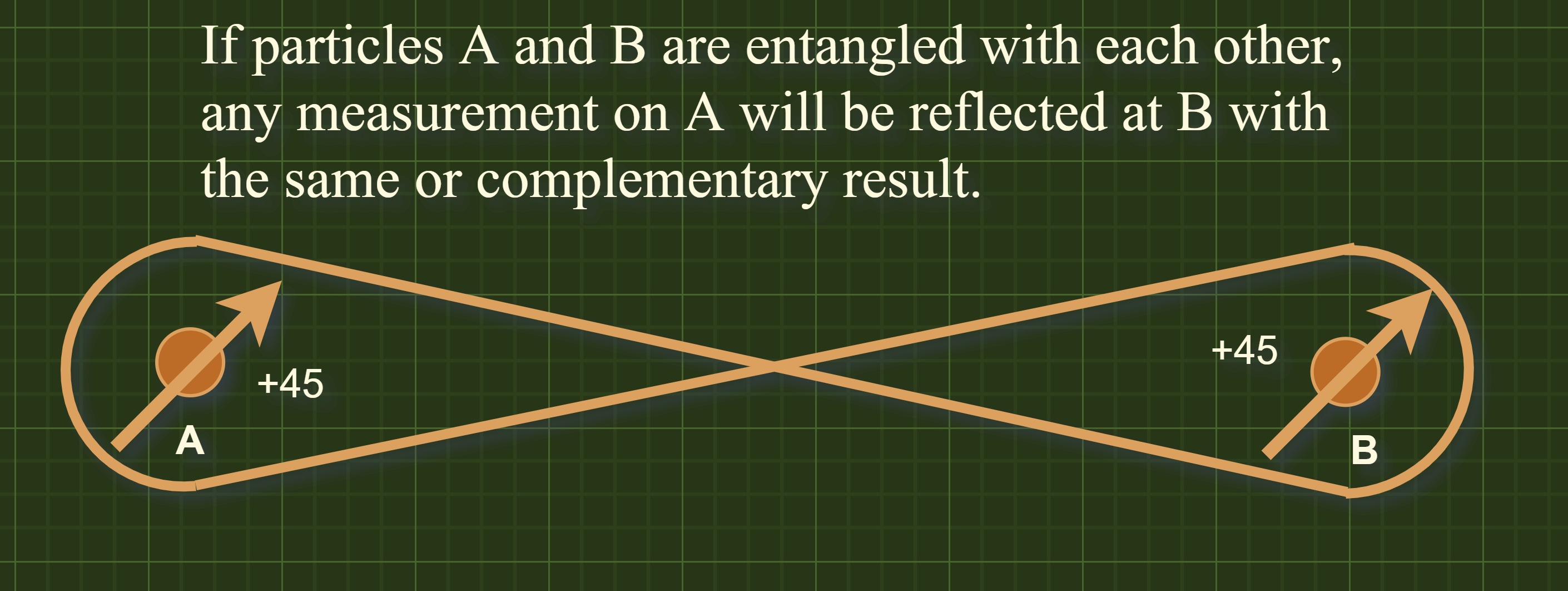}
\caption{Entanglement: Two particles \textbf{A} and \textbf{B} are correlated with each other even if they are a long distance apart. If any measurement is done at \textbf{A} having spin +45, then the same is reflected at \textbf{B}. This happens due to the weird phenomena of quantum entanglement.}
    \label{fig:Fig 5}
\end{figure}
\subsection{Multiple Qubits}
\label{sec: Sec. II-C}
In \textbf{Sec. \ref{sec: Sec. II-B}}, it was seen that a single qubit system is 2D but, in the real world, more number of qubits are required for complex computations to gain full advantage of quantum postulates. Hence, there is also a need to study multiple qubit systems. For instance, a two-qubit system will have four computational basis states denoted by $\big|00\big\rangle$, $\big|01\big\rangle$, $\big|10\big\rangle$, $\big|11\big\rangle$. 
The quantum state for a pair of qubits is represented by state vectors having their respective amplitude coefficients as
\begin{equation}\label{eq4}
    \big|\psi\big\rangle=a_{00}\big|00\big\rangle+a_{01}\big|01\big\rangle+a_{10}\big|10\big\rangle+\textcolor{black}{a_{11}}\big|11\big\rangle.
\end{equation}
Similar to the condition that has been discussed for a single qubit, the measurement result of \eqref{eq4} occurs with probability ${a_{00}^2}$ for $\big|00\big\rangle$, ${a_{01}^2}$ for $\big|01\big\rangle$, ${a_{10}^2}$ for $\big|10\big\rangle$, and ${a_{11}^2}$ for $\big|11\big\rangle$. If \textcolor{black}{\textit{x} = 00, 01, 10, 11}, then normalization condition is $\sum_x |a_{x}|^2 = 1 $. \textcolor{black}{Thus,} it can be seen that with the addition of single qubit, the computational basis states are increased exponentially thereby giving space for complex computations possible that were almost impossible with classical resources.

\subsubsection{Bell states or EPR Pairs}
\label{sec: Sec. II-C-1}
EPR is the name which came from famous paradox raised by three scientists Einstein, Podolsky, and Rosen in their famous paper \textit{"Can a quantum mechanical description of physical reality be considered complete?”}, where strange properties of Bell states were pointed out \cite{einstein1935can}. In a two qubit mechanical system, \textcolor{black}{Bell states are mathematically expressed as: \begin{equation}\label{eq5}
\big|\Phi^+\big\rangle=\frac{1}{\sqrt{2}}\big|00\big \rangle+\frac{1}{\sqrt{2}}\big|11\big \rangle,
    \end{equation}
    \begin{equation}\label{eq6}
\big|\Phi^-\big\rangle=\frac{1}{\sqrt{2}}\big|00\big \rangle-\frac{1}{\sqrt{2}}\big|11\big \rangle,
    \end{equation}
    \begin{equation}\label{eq7}
\big|\Psi^+\big\rangle=\frac{1}{\sqrt{2}}\big|01\big \rangle+\frac{1}{\sqrt{2}}\big|10\big \rangle,
    \end{equation}
    \begin{equation}\label{eq8}
\big|\Psi^-\big\rangle=\frac{1}{\sqrt{2}}\big|01\big \rangle-\frac{1}{\sqrt{2}}\big|10\big \rangle.
    \end{equation}
The two particles in these Bell states are termed \textit{EPR pairs} and are responsible for entanglement generation and distribution in the case of quantum teleportation, which is the key functionality of \textcolor{black}{QI}. \textit{EPR pairs} are generated by various schemes \textbf{(Sec. \ref{sec: Sec. III-A})} and are sent to Alice and Bob so that any measurement at Alice's qubit is the same as performed on Bob's qubit, which is justified because they share entangled qubits of an EPR pair. A Bell state has the property that when the first qubit state is measured, it results in two outcomes,  i.e. 0 with $\frac{1}{2}$ probability and 1 with $\frac{1}{2}$ probability leaving the \textcolor{black}{post-measurement state} to be $\big|00\big\rangle$ and $\big|11\big\rangle$, respectively. Due to this, if the second qubit is measured, it results in the same outcome as the first qubit. This explains the interesting \textit{correlation} between these two qubits. Even if some local operations are applied to the first or second qubit, these \textit{correlations} still exited because of entanglement. EPR’s paradox was improvised by John Bell in \cite{bell1964einstein}, where it is stated that these measurement correlations in the Bell state are much stronger than anything possible in classical systems. This makes possible a processing of the information that go well beyond what is possible in the classical world.}
    

\subsubsection{Entanglement}
\label{sec: Sec. II-W-3}
Quantum entanglement, shown in \textbf{Fig. \ref{fig:Fig 5}}, is the phenomenon of correlation between two pairs of qubits that are separated by a physical distance, such that any random measurement on entangled qubits will generate the same set of random outcomes \cite{hagley1997generation}. The concept of \textit{quantum entanglement}, which is governed by laws of quantum mechanics, is the core functionality for many applications and does not have any counterpart in \textit{classical communication}. As discussed in the \textbf{Sec. \ref{sec: Sec. II-C}-1}, an EPR pair is a maximally entangled pair of qubits that are in a superposition, with the same amount of ‘1's and ‘0's. Independently measuring these qubits results in a random distribution of ‘1' and ‘0' in equal probability. This means that the state of entangled qubits is instantaneously fixed when the other pair of an entangled qubit is measured. \textcolor{black}{Entanglement does not allow any exchange of information between remote nodes, rather it allows obtaining only mutual information. For information exchange, qubits can be transmitted between remote nodes without actually sending them from source to destination by the phenomenon of quantum \textit{teleportation} \cite{cacciapuoti2020entanglement}, \cite{bennett1993teleporting}, which is a fundamental for quantum networks \cite{Kimble2008}. An unknown quantum state can also be transmitted through quantum teleportation from source to a destination over long distances with the aid of EPR pair and local operations.}

The concept of \textit{entanglement} was first recognised by Einstein, Podolsky, and Rosen in 1935 \cite{einstein1935can} and the phenomena of entanglement was originally given by \textcolor{black}{Schrödinger} in the same year \cite{schrodinger1935gegenwartige}. It was also termed as \textit{spooky action at a distance} by Einstein. After that, \textcolor{black}{Bohm} investigated the two quantum particles, that were entangled to each other, are now termed qubits \cite{bohm1952suggested}. \textcolor{black}{Later on}, \textcolor{black}{Bell} in his research \cite{bell1964einstein} come up with a theorem to experimentally test the question of  the locality and reality of entanglement. This development motivated various researchers to create entangled particles in labs \cite{freedman1972experimental}, \cite{aspect1982experimental}, \cite{ou1988violation}. A series of experiments practised in 2015 covered all the experimental loopholes that were present before \cite{hensen2015loophole}, \cite{giustina2015significant}, \cite{shalm2015strong}. Using entanglement for communication in technical applications is not clear, because entanglement itself does not allow information transmission. However, it can help to build a virtual secure channel or a cryptographic channel \cite{ekert1991quantum}. Systems with multiple particles involved are represented by $2^{n}$-Dimensional state space. It can be seen that, with the addition of a single qubit, the computational basis states are increased exponentially, thereby giving space for complex computations which are impossible in classical communication systems. A system of this type is the Greenberger–Horne–Zeilinger (GHZ) theorem \cite{greenberger1989going}, which is different from Bell's theorem \cite{bell1964einstein} in the context of deterministic nature of the predicted outcomes. The application of the GHZ theorem is that multi-particle entanglement can be used for QEC \cite{shor1995scheme}. These ways high-dimensional entangled EPR pairs are developed and proved advantageous in quantum theory \cite{erhard2020advances}.

\begin{figure}[!t]
\centering
\includegraphics[width=0.5\textwidth]{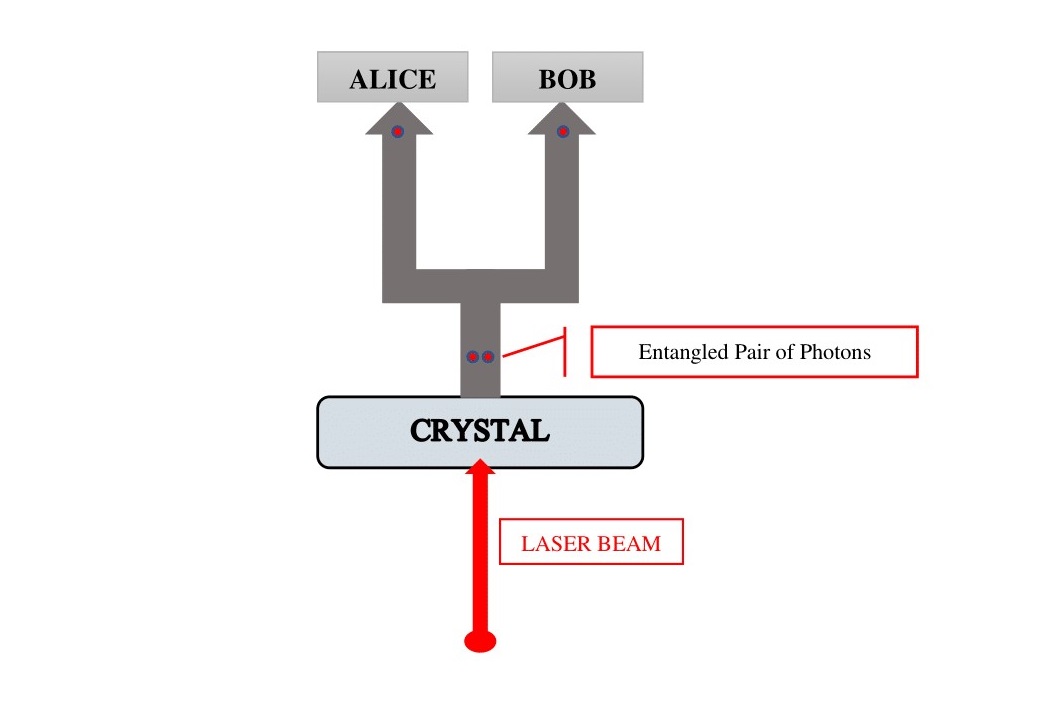}
\caption{Parametric Down Conversion: Generation of entanglement by incident of laser beam to a crystal.}
    \label{fig:Fig 6}
\end{figure}
\subsection{Quantum Gates}
\label{sec: Sec. II-D}

Quantum computations rely on quantum gates for changing or manipulating the states of qubits. A quantum computer is built using quantum circuits that contain various gates and wires to manipulate the information stored in qubits. Using these quantum gates, quantum circuits can be built that can entangle and teleport qubits from Alice to Bob. In this section, some of the famous quantum gates are discussed, which are the building blocks of various quantum circuits required to build a full-fledged quantum computer. 

\subsubsection{Single qubit quantum gates}
\label{sec: Sec. II-D-1}

Single qubit operation in quantum domain is simply defined by any rotation about the axis of the Bloch sphere as shown in \textbf{Fig. \ref{fig:Fig 4}}. These rotations can be defined by Pauli matrices \cite{van2014quantum}. Certain transforms are defined for 180$^{\circ}$ rotations and are given by:
\begin{align}\label{eq9}
Pauli-X\equiv \begin{bmatrix}
   0&1\\
   1&0 
\end{bmatrix},
\end{align}
which is represented as transformation along X-axis. Similarly, for transformations along Y-axis and Z-axis, the matrix representation is given by:
\begin{align}\label{eq10}
Pauli-Y\equiv \begin{bmatrix}
   0&-\textit{i}\\
   \textit{i}&0 ,
\end{bmatrix},
\end {align}
and 
\begin{align}\label{eq11}
Pauli-Z\equiv \begin{bmatrix}
   1&0\\
   0&-1
\end{bmatrix},
\end {align}
respectively. Equation (\ref{eq9}) is the matrix representation of quantum NOT-Gate. \textit{Pauli}-X can be manipulated with a 90$^{\circ}$ ($\pi$/2) rotation along Y-axis after 180$^{\circ}$ ($\pi$) rotation about the Z-axis yielding a Hadamard gate transform represented by the following matrix:
\begin{align}\label{eq12}
H\equiv\frac{1}{\sqrt{2}}
\begin{bmatrix}
 1&1\\
   1&-1
\end{bmatrix}.
\end {align}
\begin{table}[!t]
\caption{Truth Table for CNOT-Gate}
   \centering
\begin{tabular}{ |c|c|c|c| }
\hline
 \multicolumn{2}{|c|}{Input}&\multicolumn{2}{c|}{Output}\\
  \hline
 Control Qubit& Target Qubit& Control Qubit& Target Qubit \\
 \hline
 0 & 0 &0&0\\
 \hline
 0   & 1&0& 1\\
 \hline
 1&1&1&0\\
 \hline
 1&0&1&1\\
 \hline
\end{tabular}
\label{tab: II}
\end{table}

\subsubsection{Multiple Qubits Gate}
\label{sec: Sec. II-D-2}
In practice, more than one qubit is involved in computations. Hence, studying manipulations on multiple qubits is important to understand the insights. A \textit{controlled}-NOT (CNOT) gate is the widely used multi-qubit logic gate, which has \textit{target} qubit and \textit{control} qubit as the inputs of the CNOT gate. If the \textit{control} qubit is kept one, \textit{target} qubit is inverted at the output; if the \textit{control} qubit is kept zero, the \textit{target} qubit retains its original state. To understand the operation, a truth table is shown in \textbf{Table \ref{tab: II}} and its matrix transform is:
\begin{equation}\label{eq13}
CNOT_{Gate}\equiv
\begin{bmatrix}
 1&0&0&0\\
   0&1&0&0\\
   0&0&0&1\\
   0&0&1&0\\
\end{bmatrix}.
\end{equation}

\section{ENABLING TECHNOLOGIES}
\label{sec: Sec. III}
This section starts with the weird phenomena of quantum mechanics,  i.e. quantum entanglement, discussing the schemes for entanglement generation and distribution, followed by technologies adopted for representing qubits for quantum computation and networking. All these schemes and the choices for the selection of qubits are summarized in \textbf{Table \ref{tab: III}}. 

\subsection{Entanglement generation and distribution}
\label{sec: Sec. III-A}

Entanglement is a counter-intuitive form of correlation that has no correspondence in the classical domain as discussed earlier. Random results can be obtained by individually measuring any qubit forming an EPR pair and the two independent measurements result in an outcome that is either directly or complementary related. In particular, for exchange of information between two remote nodes through quantum teleportation, entanglement generation and distribution are the key ingredient in this process \cite{stenholm1998teleportation}. Some schemes adopted for generation and distribution of EPR pairs between Alice and Bob, which are long distance apart, are discussed in what follows. 

\subsubsection{Spontaneous Parametric Down Conversion}
\label{sec: Sec. III-A-1}
In this scheme, which is depicted in \textbf{Fig. \ref{fig:Fig 6}}, photons are employed for entanglement generation and distribution between two remote nodes by the incidence of the laser beam on a crystal. The use of photons is advantageous as they can travel at high speed and can cover a long distance in fiber channels with the least amount of attenuation. Thus, photons are promising candidates for the entanglement creation \cite{cabrillo1999creation}, \cite{bose1999proposal}. The photons are also associated with moderate amount of decoherence and noise as they least interact with the environment \cite{cacciapuoti2020entanglement}. In the optical parametric down conversion method, a laser beam is incident on a birefringent crystal, which  converts photon beams into pair of correlated photons with the help of non-linear effects of crystal\cite{rubin1994theory}. The entangled photons are vertically and horizontally polarized and individually transported to Alice and Bob through a quantum channel, so that any measurement at Alice is reflected at Bob irrespective of the distance between them.

\subsubsection{Single Atom Excitation by Laser Beam}
\label{sec: Sec. III-A-2}
\begin{figure}[!t]
\centering
\includegraphics[width=0.5\textwidth]{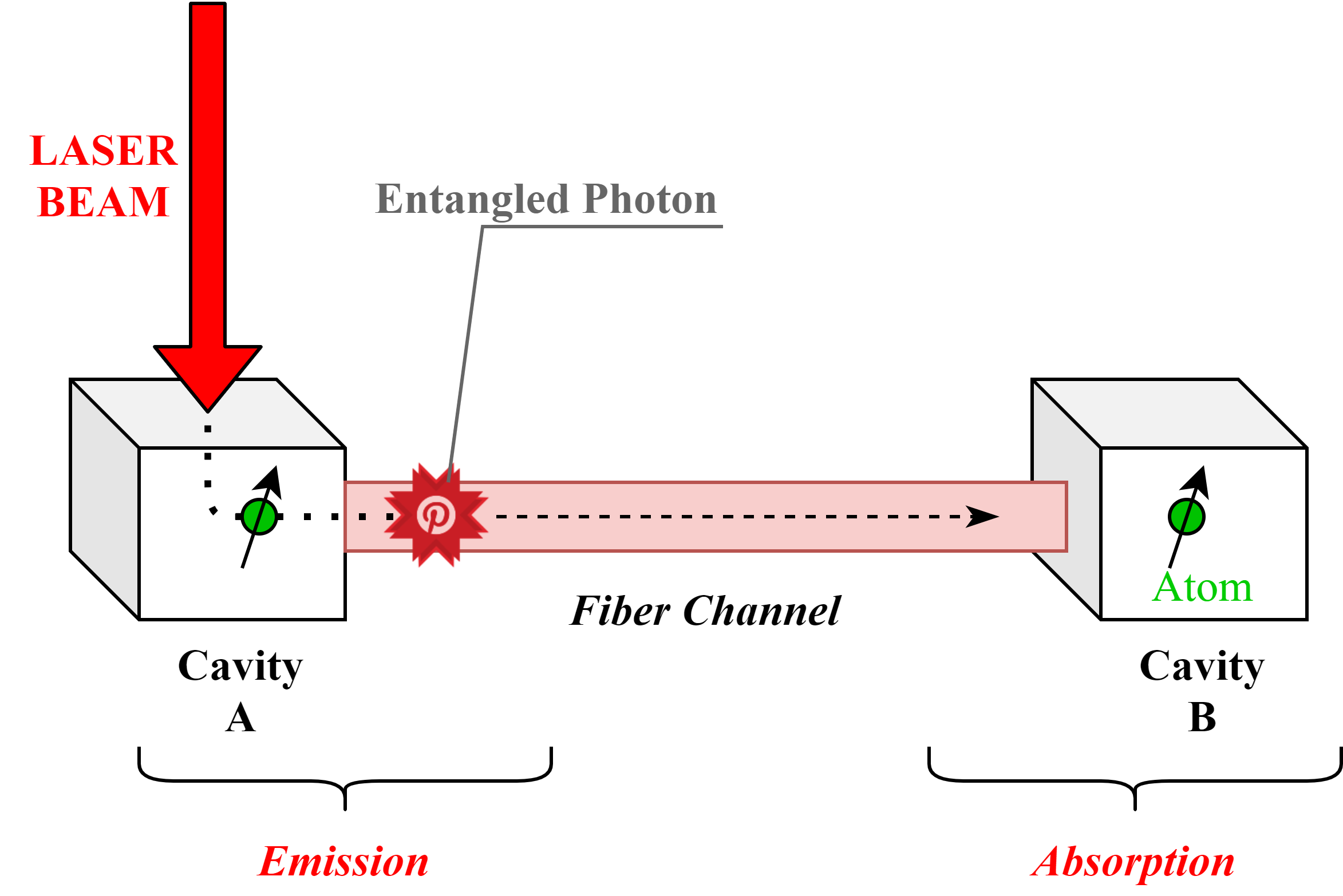}
\caption{Entanglement generation by single atom excitation.}
    \label{fig:Fig 7}
\end{figure}

 A different scheme, as shown in \textbf{Fig. \ref{fig:Fig 7}}, can be employed for entanglement generation and distribution. In this, atoms coupled firmly with an optical cavity are excited by a laser source, resulting in the radiation of atom-entangled photons that departs from the cavity $C_{A}$ and travels through quantum channel reaching another cavity $C_{B}$, where the photons are absorbed coherently, thus mapping the polarization of photon onto the state of remote atoms. In this way, two atoms are remotely entangled by photons \cite{ritter2012elementary}, \cite{welte2018photon}. 

\subsubsection{Two Atoms simultaneous Excitation by Laser Beam}
\label{sec: Sec. III-A-3}
\begin{figure}[!t]
\centering
\includegraphics[width=0.5\textwidth]{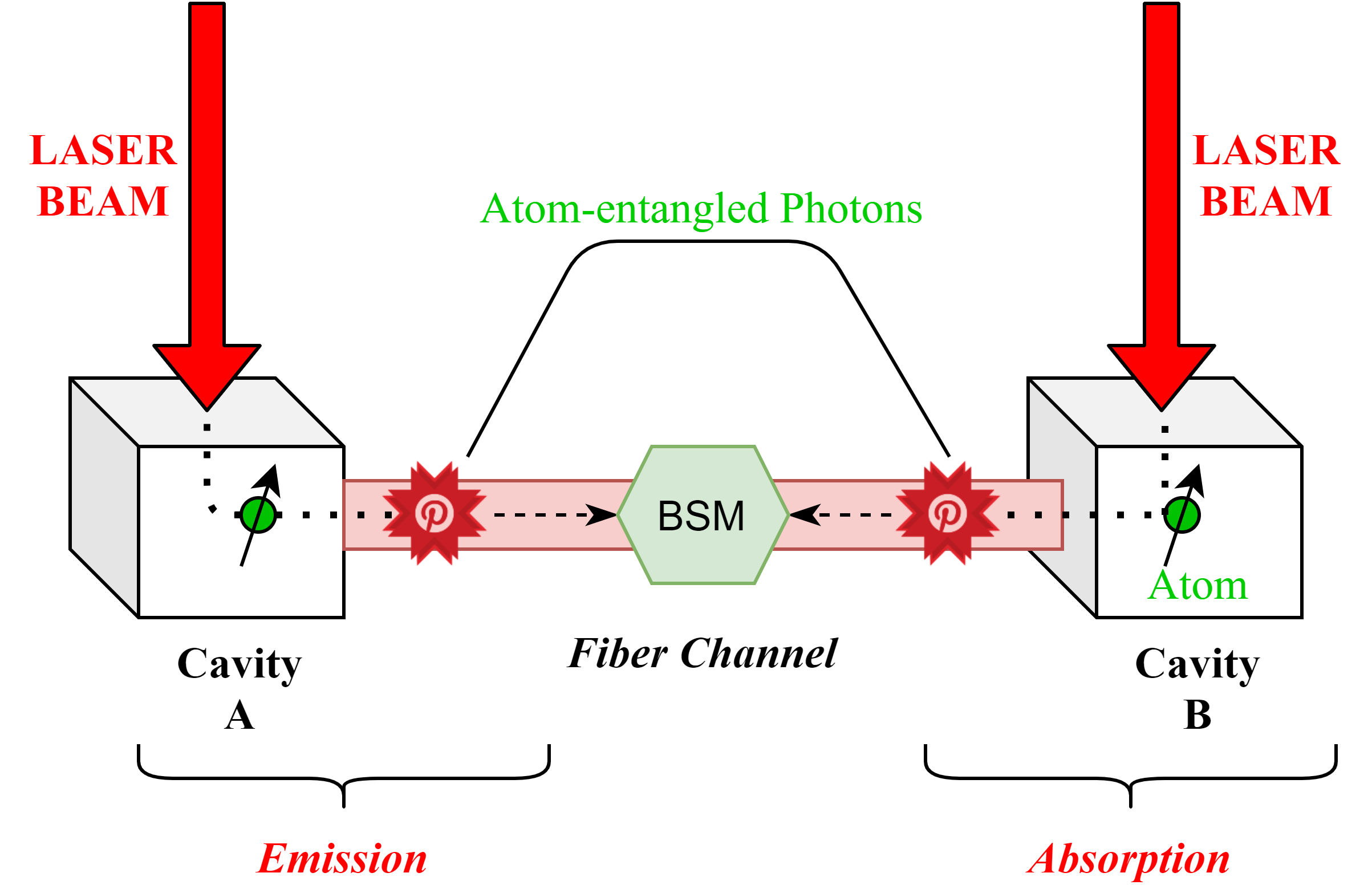}
\caption{Entanglement generation by two atoms simultaneous excitation by laser pulse.}
    \label{fig:Fig 8}
\end{figure}

 In this scheme, two atoms are simultaneously excited by the incidence of a laser beam on cavities $C_{A}$ and $C_{B}$ resulting in radiation of two atom-entangled photons. These atom-entangled photons depart from $C_{A}$ and $C_{B}$, as shown in \textbf{Fig. \ref{fig:Fig 8}}, and travel along the quantum channel where the \textcolor{black}{Bell State Measurement (BSM) takes place. BSM is a scheme that is utilized for the generation of maximally entangled Bell states based on Hong-Ou-Mandel (HOM) interference, where two photons interact with each other at a beamsplitter (BS) \cite{hong1987measurement}. It performs a joint quantum measurement on the state of these atom-entangled photons resulting in entanglement swapping operation \cite{zukowski1993event}, discussed in \textbf{Sec. \ref{sec: Sec. IV-B}}, distributing entanglement between remote atoms \cite{ferrari2010entanglement}, \cite{northup2014quantum}, \cite{caleffi2017optimal}.}
 
\subsection{Qubit Technologies}
\label{sec: Sec. III-B}

\subsubsection{Cavity quantum electrodynamics}
\label{sec: Sec. III-B-1}

Cavity quantum electrodynamics (cQED) refers to qubit technology where coherent interaction takes place between matter qubits such as quantum dot systems, trapped ions, and quantized field of optical or microwave resonator. Coherency is achieved by using a low-loss cavity, which is used to enhance the electric field associated with a single photon, such that the rabi frequency corresponding to atom-field interaction is faster than the decay rate of the field in the cavity \cite{Kimble2008}. It investigates the coupling properties of atoms with photon modes in cavities. Schwab \textit{et. al.} \cite{armour2002quantum}, \cite{irish2003quantum} have recently proposed the idea of cQED using nano-mechanical resonators \cite{armour2002quantum}, \cite{irish2003quantum}. The cQED has many applications in Quantum Information Processing (QIP) such as atom-atom, atom-photon, and photon-photon entanglement \cite{pellizzari1995decoherence}, \cite{van1997purifying}. 

\begin{table*}[ht!]
\centering
\caption{Choices for qubit technology selection}
\begin{tabular}[c]{@{}lccccccl@{}}
\toprule \toprule 
\textbf{Design questions} & \multicolumn{6}{c}{\textbf{Design choices}}\\ 
\midrule

 &\multicolumn{6}{c}{Entanglement generation}                                                         \\ \midrule
\textbf{Entities in entanglement} &
  \multicolumn{2}{c}{\begin{tabular}[c]{@{}c@{}}Photon-Photon: \\ Spontaneous Parametric Down Conversion \end{tabular}} &
  \multicolumn{2}{c}{\begin{tabular}[c]{@{}c@{}}Photon-Atom:\\ Single atom excitation\end{tabular}} &
  \multicolumn{2}{c}{\begin{tabular}[c]{@{}c@{}}Atom-Atom:\\ Two atoms excitation\end{tabular}} \\ \midrule
\textbf{Technical overhead vs. success rate} &
  \multicolumn{3}{c}{\begin{tabular}[c]{@{}c@{}}more overhead, high success rate: \\ deterministic interface\end{tabular}} &
  \multicolumn{3}{c}{\begin{tabular}[c]{@{}c@{}}less overhead, non-deterministic:\\ heralded interface\end{tabular}} \\ \midrule\midrule
 \textbf{Features} &\multicolumn{6}{c}{\textbf{Qubit technologies}}                                                         \\ \midrule
Reusing existing semiconductor production & \multicolumn{6}{c}{Solid State Approach}     \\
Cheap production at current scale               & \multicolumn{6}{c}{Superconducting approach} \\
Low cost of running                             & \multicolumn{6}{c}{Photonic approach}        \\
High connectivity                               & \multicolumn{6}{c}{Ion trap approach}        \\ \bottomrule \bottomrule
\end{tabular}
\label{tab: III}
\end{table*}

\subsubsection{Photonic qubits}
\label{sec: Sec. III-B-2}
Photonic qubits have played a very important role in QIP as they can be controlled by traditional optical components and experience the least amount of \textit{decoherence} in their path, which is the phenomenon of loss of quantum information due to the fragile nature of photons in noisy environments. As it is one of the most challenging and disrupting aspect of quantum communication, it will be detailed in \textbf{Sec. \ref{sec: Sec. VI-A}}. This is the first qubit technology that was realised for generation and distribution of quantum entanglement \cite{ou1988violation}, \cite{shih1988new}, \cite{kwiat1995new} and has experimentally realized quantum cryptography \cite{gisin2002quantum}. They are the first candidates to accomplish the task of quantum teleportation \cite{pirandola2015advances}, \cite{bouwmeester1997experimental}, \cite{pan1998experimental}, \cite{pan2001experimental}, \cite{boschi1998experimental}, \cite{furusawa1998unconditional}. As \textcolor{black}{QI} requires long-distance communication, photonic qubits can achieve this with very little decoherence. If free space is considered as a communicating channel, the birefringence is weak and the absorption of photons is small at optical frequencies as compared to fiber channels \cite{zhong2020quantum}. Also, this technology is particularly attractive as photons can be instantly connected to various quantum communication applications such as distributed quantum computation \cite{cuomo2020towards}.

\subsubsection{Trapped ions}
\label{sec: Sec. III-B-3}

The trapped ion is also a promising approach for scalable QIP \cite{nigg2014quantum}, \cite{gale2020optimized}. It offers excellent features such as coherence \cite{monroe2013scaling} and effective implementation of entangling gates with least amount of cross-talk \cite{lu2015dark}. This approach depends basically on the idea proposed by Cirac and Zoller in \cite{cirac1995quantum}. Some experiments using trapped-ions qubit show single-qubit and multi-qubit operations having fidelity significantly higher than the minimum threshold needed during fault-tolerant quantum computing \cite{harty2014high}, \cite{gaebler2016high}, \cite{ballance2016high}. Using this technique, a fully programmable five-qubit quantum computer was built \cite{debnath2016demonstration}. Although the ion trap approach can hold a large number of qubits, their role in quantum computing is radically restricted by the number of entanglement operations that can be executed within the time limits of coherence \cite{divincenzo1995quantum}. Consequently, the improvement of mechanisms for achieving high-speed and high-fidelity entangling gates is necessary for accomplishing large-scale computation on trapped ion platforms \cite{ladd2010quantum}.

\subsubsection{Solid state approach}
\label{sec: Sec. III-B-4}
Solid-state qubits are formed by electron spins \cite{loss1998quantum} followed by superconducting circuits \cite{kjaergaard2020superconducting} having the transition frequencies in the range of Gigahertz (GHz) thus making them compatible with traditional off-the-shelf Radio Frequency (RF) and microwave components. The solid-state quantum memories, along with photonic qubits, provide excellent solutions that comprise semiconductor quantum dots \cite{loss1998quantum}, \cite{gao2013quantum} or rare-earth-doped crystals \cite{bussieres2014quantum}. This approach has the advantage of meeting scalability requirements in QIP. A large number of solid-state systems has been built, starting from the 1990s up to recently, that can achieve quantum computation with required scalability. Big tech giants like Google, Intel, and IBM are spending profoundly on this technology and due to the demonstration of the best performance of superconducting qubits, they have already invested into building supreme quantum processors. The most extensive demonstration is the Sycamore processor, built by Google, that has 53 operational qubits \cite{arute2019quantum} and is considered as a breakthrough in the development of technology to realize \textcolor{black}{QI} on a large scale \cite{oliver2019quantum}. 
 
\subsubsection{Superconducting approach}
\label{sec: Sec. III-B-5}

Superconducting circuits comprises Josephson junctions, interconnects, and passive elements, i.e. inductors and capacitors, that provides good isolation \cite{krantz2019quantum}. These circuits operate at milliKelvin (mK) temperatures. Some materials used in such circuits are aluminium (Al), titanium nitride (TiN), and niobium (Nb), substrates of silicon (Si) and sapphire ($Al_{2}O_{3}$), which are compatible with CMOS manufacturing technology \cite{yost2020solid}. This approach provides lithographic scalability, compatibility with off-the-shelf microwave components, and nanosecond-level operability. These features have all been fused, thus placing superconducting qubit technology at the forefront of the development of emerging quantum processors. This technology also features high fidelity teleportation \cite{steffen2013deterministic} with single-qubit gate fidelities exceeding 99.9\% and two-qubit gate fidelities exceeding 99.5\% \cite{kjaergaard2020superconducting}.

\section{QUANTUM INTERNET FUNCTIONALITIES}
\label{sec: Sec. IV}
Quantum Internet is a system that consists of various functionalities required for communicating information among remote nodes of a network. In this section, we will focus on various functionalities of QI like quantum teleportation, quantum channels, quantum repeaters, quantum memories, QKD, and end nodes. 

\subsection{Quantum Teleportation}
\label{sec: Sec. IV-A}
Quantum teleportation comprises a bewildering system for transmitting qubits inside a network of quantum communication without physically transferring the particle storing the qubit. The concept of teleporting qubits has started with the laid work \textit{“Teleporting an  unknown quantum state via dual classical and \textcolor{black}{Einstein-Podolsky-Rosen channels}”} \cite{bennett1993teleporting}. \textcolor{black}{It was proved experimentally by Anton Zeilinger group in 1997 for the first time through the successful transmission of quantum optical states \cite{bouwmeester1997experimental}. At the same time, another group of Sandu Popescu successfully experimented the photonic teleportation in 1998 \cite{boschi1998experimental}. The teleportation distance of 55 m was achieved under laboratory conditions by Swiss researcher Nicolas Gisin and his team in 2003 \cite{marcikic2003long} and later verified it in 2007 by using Swisscom's commercial telecommunication network based on fiber-optics \cite{landry2007quantum}. Later in 2004, researchers from Innsbruck and USA succeeded in teleporting atomic states for the first time \cite{barrett2004deterministic} followed by an Austrian group led by Anton Zeilinger who achieved a distance of 600 m through a fiber-optic across Danube \cite{ursin2004quantum}. This distance was improved to 1.3 km in \cite{hensen2015loophole}. Significant experimental progresses in quantum teleportation networks have improved the beeline distance to 8.2 km \cite{valivarthi2016quantum} and 100 Km \cite{yin2012quantum}. A larger distance - 144 km between the islands of La Palma and Tenerife - of teleportation was further achieved by a team of Anton Zeilinger in 2012 using a free space channel \cite{ursin2007entanglement}. With the advancements in satellite technology, teleportation over a distance of more than 1200 km has been successfully realized with the launch of the Micius satellite by Jian-Wei Pan and his team from China \cite{ren2017ground}.} 

\begin{figure}[!t]
\centering
\includegraphics[width=0.55\textwidth]{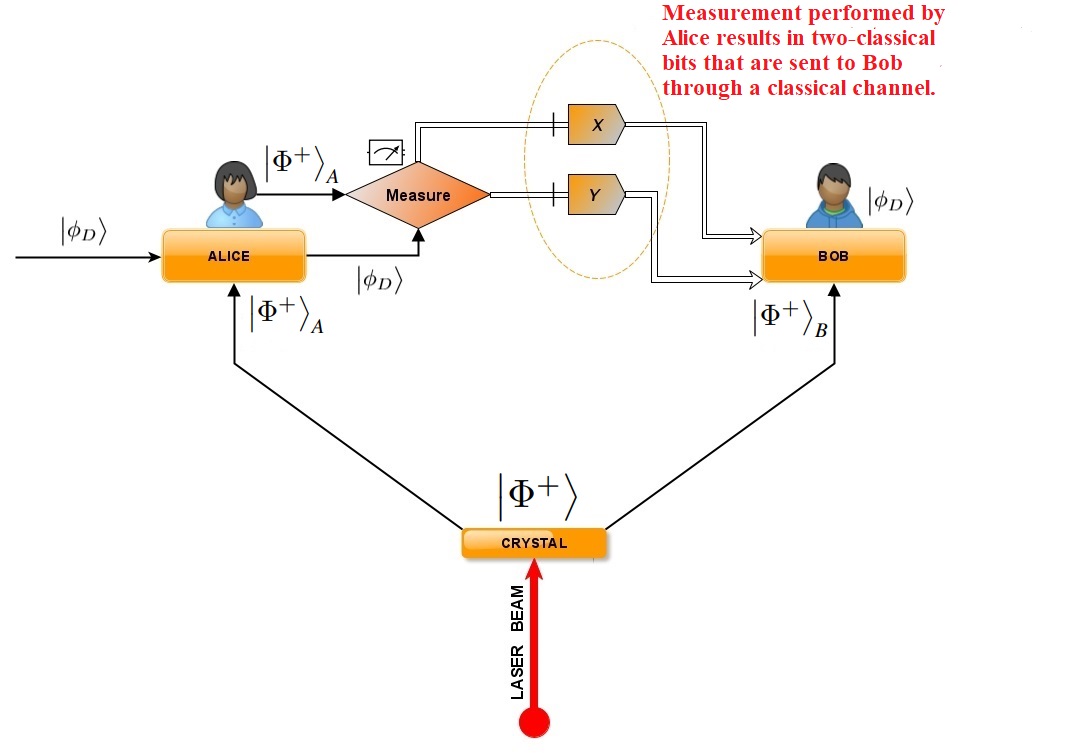}
\caption{Quantum Teleportation: Alice holds $\big|+\big\rangle_{\textit{A}}$ and Bob holds $\big|+\big\rangle_{\textit{B}}$ state of the pair. Measurement of Bell state and data qubit as a pair $\big|\phi_D\big\rangle$$\big|+\big\rangle_{\textit{A}}$ is done by Alice and the result of the measurement is stored in classical bits \textbf{X} and \textbf{Y} as shown above. Manipulating \textbf{X} and \textbf{Y} at Bob, the original state of \textbf{$\big|\phi_D\big\rangle$} is recovered.}
    \label{fig:Fig 9}
\end{figure}

A large portion of the examinations has been conﬁned to the teleportation of single-body quantum states,  i.e. quantum teleportation of two-level states \cite{bennett1993teleporting}, multidimensional states \cite{stenholm1998teleportation}, continuous variable teleportation \cite{braunstein1998teleportation}, and discrete variable teleportation \cite{gorbachev2000quantum}. Teleportation through noisy channels of bipartite entangled states is studied in \cite{lee2000entanglement}. Quantum teleportation of a two-qubit entangled state is discussed in \cite{ikram2000quantum} and \cite{shi2000probabilistic}. 
It is facilitated by a strange phenomenon of quantum mechanics called \textit{quantum entanglement}. Quantum teleportation is realized, as shown in \textbf{Fig. \ref{fig:Fig 9}}, with two parallel links required for communication:
\begin{enumerate}
  \item Quantum channel link, for generating entangled EPR pair and distributing it between source and destination \textcolor{black}{nodes.} 
  \item \textcolor{black}{Classical channel link, for transmitting two classical bits from a source node to a destination node for sending the measurement result of source qubit.}
\end{enumerate}
Quantum teleportation is quite intimidating when it comes to its application, which renders efficient communication and fidelity of the system \cite{cacciapuoti2020entanglement}. However, at the same time, it involves imperfections in its process imposed by photon loss in environment-\textit{decoherence} or are the result of a sequence of operations that are applied for the processing of teleportation \cite{he2006influence}. Thus, the technology opted for quantum teleportation plays a strong part in improving the system fidelity and reducing decoherence \cite{cacciapuoti2020entanglement}. 

The process of teleportation begins with source node Alice in possession of an unknown data qubit $\big|\phi_D\big\rangle$ = $\alpha\big|0\big\rangle + \beta\big|1\big\rangle$, and an EPR pair shared between Alice and Bob. Let us assume the bell pair to be $\big|\Phi^+\big\rangle$, as shown in (\ref{eq5}). \textcolor{black}{Bell pair qubit} which Alice holds is denoted by $\big|\Phi^+\big\rangle_{\textit{A}}$ and the other pair of qubit held by Bob is $\big|\Phi^+\big\rangle_{\textit{B}}$, as shown in \textbf{Fig. \ref{fig:Fig 9}}. The procedure of teleportation between Alice and Bob is as follows:
\begin{enumerate}
 \item Bell state $\big|\Phi^+\big\rangle$ is prepared by Alice and Bob such that each of them holds the qubit of this EPR pair. Alice holds $\big|\Phi^+\big\rangle_{\textit{A}}$ and Bob holds $\big|\Phi^+\big\rangle_{\textit{B}}$ state of the pair.
  \item Simultaneously, the state to be teleported,  i.e.  $\big|\phi_D\big\rangle$, is acquired by Alice.
   \item Measurement of Bell state and data qubit as a pair $\big|\phi_D\big\rangle \big|\Phi^+\big\rangle_{\textit{A}}$ is performed by Alice and the result of measurement, stored in two classical bits, is sent to destination Bob through classical link.
    \item Bob manipulates the received classical bits and apply corrections in X or Z direction accordingly.
    \item At Last, Bob's qubit state  $\big|\Phi^+\big\rangle_{\textit{B}}$ now holds the same quantum state as Alice's $\big|\phi_D\big\rangle$. This completes the process of teleportation of the data qubit between end nodes of Alice and Bob.
    \end{enumerate}
    
In the process of measurement, the entanglement of EPR pair is destroyed, therefore it should be noted here that for teleportation of another data qubit, another EPR pair should be generated and distributed. Thus, it can be seen that quantum entanglement is an important quantum effect, as it is fundamental pre-requisite for transmission of particle storing the qubit. It is vital to look at the technologies to generate and distribute the qubits, which are discussed in \textbf{Sec. \ref{sec: Sec. III-B}}. In view that Alice and Bob represent remote nodes, the entanglement generation taking place at one side must be complemented by way of the entanglement distribution capability, moving entangled qubit to the destination side. Considering this, it is highly likely for the adoption of photons as entanglement providers \cite{northup2014quantum}. The reason for this choice lays in the advantages offered by photons for entanglement distribution \cite{kwiat1995new}, along with least interaction with the environment, smooth control with basic optical components as well as high-speed low-loss transmission to far off nodes.

\subsubsection{Mathematical details}
\label{sec: Sec. IV-A-1}
For the state of clarity, lets analyse the mathematical details of quantum teleportation process. 
The input is data qubit state $\big|\phi_D\big\rangle$, that is to be teleported from Alice to Bob. EPR pairs of Bell state $\big|\Phi^+\big\rangle$ is required to be in possession of both Alice and Bob.

\begin{itemize}
    \item Global state of Alice and Bob is represented by $\big|\psi_D\big\rangle$, which is the tensor product of $\big|\phi_D\big\rangle$ and $\big|\Phi^+\big\rangle$ given by:
    \begin{equation}{\label{eq14}}
        \big|\psi_D\big\rangle=\big|\phi_D\big\rangle\otimes\big|+\big\rangle.
    \end{equation}
Thus, we get
    \begin{equation}{\label{eq15}}
\big|\psi_D\big\rangle=\alpha\big|0\big\rangle\otimes\frac{\big|00\big \rangle+\big|11\big\rangle}{\sqrt{2}}+\textcolor{black}{\beta\big|1\big\rangle}\otimes\frac{\big|00\big \rangle+\big|11\big\rangle}{\sqrt{2}},
    \end{equation}
Now, (15) becomes    
    \begin{equation}{\label{eq16}}
    \big|\psi_D\big\rangle=\frac{(\alpha\big|000\big\rangle+\alpha\big|011\big\rangle+\beta\big|100\big\rangle+\textcolor{black}{\beta}\big|111\big\rangle)}{\sqrt{2}}.
    \end{equation}
    
    \item Alice then applies the CNOT operation to the pair of qubits that she has, that maps $\big|10\big \rangle$ into $\big|11\big \rangle$ and vice versa, the result of which is:
    \begin{equation}{\label{eq17}}
       \big|\psi_D\big\rangle=\frac{(\alpha\big|000\big\rangle+\alpha\big|011\big\rangle+\beta\big|110\big\rangle+\textcolor{black}{\beta}\big|101\big\rangle)}{\sqrt{2}}. 
    \end{equation}
    
    \item Alice then applies Hadamard gate operation to $\big|\phi_D\big\rangle$, that maps $\big|0\big\rangle$ into $\frac{\big|0\big\rangle+\big|1\big\rangle}{\sqrt{2}}$, and $\big|1\big\rangle$ into $\frac{\big|0\big\rangle-\big|1\big\rangle}{\sqrt{2}}$, resulting in:
    \begin{align}{\label{eq18}}
                \big|\psi_D\big\rangle=\big(\alpha\big|000\big\rangle+\alpha\big|100\big\rangle+\alpha\big|011\big\rangle+\alpha\big|111\big\rangle\nonumber\\+\beta\big|010\big\rangle-\beta\big|110\big\rangle+\beta\big|001\big\rangle-\beta\big|101\big\rangle\big)/2.     
        \end{align}
        
    \item Rearranging (18) gives 
    \begin{align}\label{eq19}
        \big|\psi_D\big\rangle=\Big(\big|00\big\rangle\otimes\big(\alpha\big|0\big\rangle+\beta\big|1\big\rangle\big)+\big|01\big\rangle\otimes\big(\alpha\big|1\big\rangle+\beta\big|0\big\rangle\big)\nonumber\\+\big|10\big\rangle\otimes\big(\alpha\big|0\big\rangle-\beta\big|1\big\rangle\big)+\big|11\big\rangle\otimes\big(\alpha\big|1\big\rangle-\beta\big|0\big\rangle\big)\Big)\big/2.
    \end{align}
    
    \item Alice then applies the join measurement on the pair of qubits at her possession and has 25\% probability of getting each of the combinations $\big|00\big\rangle$, $\big|01\big\rangle$, $\big|10\big\rangle$, $\big|11\big\rangle$.
    
    \item The measurement result of Alice instantaneously affects the state of the Bob's qubit regardless of the distance between them as the pairs are entangled.
    
    \item Bob can recover the original state of qubit $ \big|\Phi_D\big\rangle$ after manipulating the classical bits, received through classical link, having the information of Alice's measurement.
    
    \item If the result of Alice's measurement is $\big|00\big\rangle$, then the state of Bob's qubit is $\alpha\big|0\big\rangle+\beta\big|1\big\rangle$ as depicted from \eqref{eq19}. Thus, Bob's qubit follows the original quantum state $\big|\Phi_D\big\rangle$ and there is no need for any further operation. If Alice's joint measurement is $\big|01\big\rangle$, the measurement result of Bob is $\alpha\big|1\big\rangle+\beta\big|0\big\rangle$. So, Bob recovers the original quantum state $\big|\Phi_D\big\rangle$ = $\alpha\big|0\big\rangle+\beta\big|1\big\rangle$ by applying Pauli-X gate operation. Similarly, for Alice's measurement of $\big|10\big\rangle$ and $\big|11\big\rangle$, Bob again recovers the original state by applying certain gate operations, which successfully completes the process of teleportation of qubits between Alice and Bob. 
    
    \item After all this, it can be concluded that teleportation process obeys the speed of light, thus preserving the theory of relativity.
\end{itemize}

\subsubsection{Teleportation using Quantum Switch}
\label{sec: Sec. IV-A-2}
Quantum teleportation, as discussed earlier, is the core functionality of \textcolor{black}{QI} that exploits the unmatched property of quantum mechanics,  i.e. \textit{quantum entanglement}. Entanglement generation and distribution is accomplished between Alice and Bob through quantum and classical channels. Since entanglement distribution is limited by noise and decoherence, it further degrades the efficiency of teleported information. Hence, there is a need to efficiently use the resources that are available by exploiting the laws of quantum mechanics and distribute the entangled information through noisy channels with the help of superposition of causal orders of available channels \cite{salek2018quantum}. 
This is achieved through \textit{quantum switch} \cite{chiribella2013quantum}, which uses superposition principle and permits quantum particles to propagate concurrently among more than one space-time trajectories \cite{caleffi2020quantum}, \cite{chiribella2019quantum}.
The usual assumption is that quantum information, which is stored in quantum particles is sent through well defined orders of channels as described by the classical theory of Shannon in \cite{shannon1948mathematical}. 
But, the particles can traverse through superposition of multiple channels at the same time, 
which means that the relative order is indefinite \cite{chiribella2012perfect}.
The quantum switch is a novel device that controls the order in which channels are traversed with the help of a \textit{control qubit}. This new resource can be advantageous to problems faced in fields of quantum computation \cite{feix2015quantum}, QIP \cite{chiribella2013quantum}, non-local games \cite{oreshkov2012quantum}, and communication complexity \cite{guerin2016exponential}. 
\begin{figure}[!t]
\centering
\includegraphics[width=0.57\textwidth,height=5.3cm]{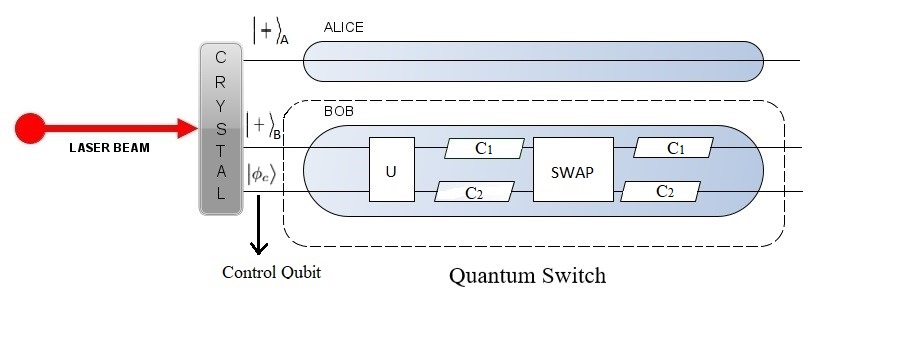}
\caption{Entanglement Distribution using quantum switch: Alice holds $\big|+\big\rangle_{\textit{A}}$ and Bob holds $\big|+\big\rangle_{\textit{B}}$ state of an EPR pair. A control qubit $\big|\phi_{c}\big\rangle$ is also there at Bob which decides the order of channel to be traversed. This operation is performed by \textit{U} gate, \textcolor{black}{that directs qubit $\big|+\big\rangle_{\textit{B}}$ at Bob through upper or lower line on the basis of the state of control qubit $\big|\phi_{c}\big\rangle$. SWAP gate enables the entanglement-carrier $\big|+\big\rangle_{\textit{B}}$ through it to the other side of the switch, distributing entanglement between Alice and Bob.}}
    \label{fig:Fig 10}
\end{figure}
Let us assume that we have two channels in the entanglement distribution process,  i.e. $C_{1}$ and $C_{2}$. If a message \textit{‘M'} travels from Alice to Bob through these channels, it can traverse either $C_{1}$ followed by $C_{2}$, or $C_{2}$ followed by $C_{1}$. But, knowing the property of quantum superposition of channels, it is not possible to determine the order in which the channel is traversed, which means that superposition of both the channels exist,  i.e. message \textit{‘M'} traverses the channel $C_{1}$ and $C_{2}$ in superposition of orders $C_{1}$$\rightarrow$$C_{2}$ and $C_{2}$$\rightarrow$$C_{1}$. For this task, a control bit $\big|\phi_{c}\big\rangle$ is assigned, as shown in \textbf{Fig. \ref{fig:Fig 10}}. If $\big|\phi_{c}\big\rangle$ is initialized to $\big|\phi_{c}\big\rangle$ = $\big|0\big\rangle$, the quantum switch enables message ‘M' to traverse path $C_{1}$$\rightarrow$$C_{2}$ and if $\big|\phi_{c}\big\rangle$ is initialized to $\big|\phi_{c}\big\rangle$ = $\big|1\big\rangle$, the quantum switch enables message \textit{‘M'} to traverse path $C_{2}$$\rightarrow$$C_{1}$. 
Interestingly, if control bit is initialized with superposition of basis states $\big|0\big\rangle$ and $\big|1\big\rangle$ in state $\big|+\big\rangle=\frac{1}{\sqrt{2}}\big|0\big\rangle+\frac{1}{\sqrt{2}}\big|1\big \rangle$, then quantum switch enables the message \textit{‘M'} to traverse the superposition of orders $C_{1}$$\rightarrow$$C_{2}$ and $C_{2}$$\rightarrow$$C_{1}$. 
\textcolor{black}{This operation of traversing the particular order of channel is performed by a \textit{U} gate as seen in \textbf{Fig. \ref{fig:Fig 10}}, directing qubit $\big|+\big\rangle_{\textit{B}}$ at Bob on the basis of state of control qubit $\big|\phi_{c}\big\rangle$ as mentioned earlier. The SWAP gate is implemented with Polarization Beam Splitter (PBS) and Half-Wave
Plates (HWPs), which helps to switch between the polarization states of $\big|\phi_{c}\big\rangle$, i.e. horizontal and vertical polarization.} 
The theoretical analysis of employing a quantum switch and advantages it offers, is discussed in \cite{salek2018quantum}, \cite{chiribella2019quantum}, \cite{guerin2019communication}, \cite{chiribella2021indefinite}, and experimentally detailed in \cite{goswami2018indefinite}, \cite{guo2020experimental}. Therefore, a quantum switch is the core functionality of the quantum web and is a key fixing in the quantum teleportation process.


\subsection{Quantum Repeaters}
\label{sec: Sec. IV-B}
\begin{figure}[!t]
\centering
\includegraphics[width=0.5\textwidth,,height=7cm]{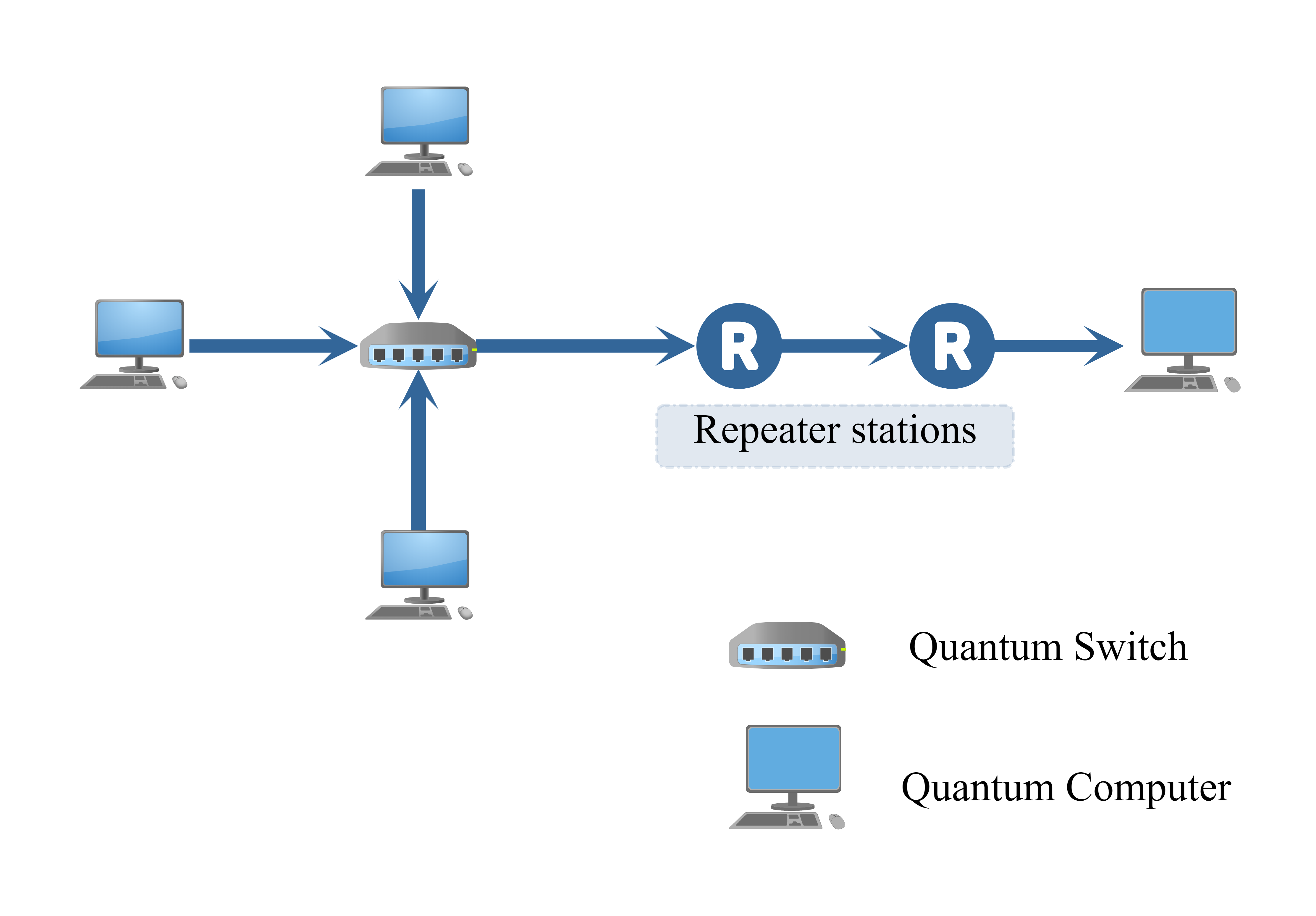}
\caption{Repeater Stations: \textcolor{black}{These stations are employed at short distances so that entanglement can be established between intermediary nodes and ultimately between end nodes which are long distance apart. Here, \textit{quantum switch} is also employed to transmit information between end nodes, traversing channels in superposition of different orders.}}
    \label{fig:Fig 11}
\end{figure}

A quantum repeater is another functionality of QI that is employed in between the remote nodes, to transmit quantum information to long distance as shown in \textbf{Fig. \ref{fig:Fig 11}}. They generate maximally entangled pairs between two far away end nodes by bifurcating the network into segments, establishing the long distance entanglement between end points. In long distance communication, most of the information carrying photons gets lost because of channel attenuation or other environmental factors. This exponential problem of photon loss was addressed by Jurgen Briegel in 1998 by introducing the model of a quantum repeater \cite{briegel1998quantum}.
In the classical domain, the issue is resolved by employing classical repeaters that copy the information after regular intervals for amplification. However, in the case of quantum communication, due to the no-cloning theorem \cite{dieks1982communication}, information carried in qubits cannot be copied or amplified since classical repeaters are not suitable for this communication. Hence, the quantum repeater model proposed by Jurgen Briegel is employed to transmit qubits from Alice to Bob without any loss of information \cite{briegel1998quantum}.

\textcolor{black}{The evolution of quantum repeaters is classified into three distinct generations \cite{munro2015inside}. In the \textbf{first generation} of quantum repeaters, intermediate stations are placed as shown in \textbf{Fig. \ref{fig:Fig 11}}, in such a way that links are short distance apart and probability for establishing entanglement among repeater stations is high. Due to a remarkable feature of \textit{entanglement swapping} \cite{munro2015inside}, discussed in \textbf{Sec. \ref{sec: Sec. IV-B}-2}, all these stations are interconnected in terms of entanglement from Alice to Bob \cite{bennett1993teleporting}, \cite{zukowski1993event}, thus entangling the far away end nodes.}
Each repeater station must be capable of receiving, processing, and transmitting classical and quantum information. For this, some specific protocols are employed, considering the following points:
\begin{itemize}
\item Firstly, entanglement must be established between adjacent nodes.
\item Next, \textit{entanglement swapping} must take place at the repeater stations such that qubits at the end nodes are instantaneously entangled.
\item \textcolor{black}{After that, error correction takes place, where few strongly entangled states are generated from weakly entangled states by a process known as \textit{entanglement distillation} or \textit{purification} \cite{bennett1996purification}. Multiple copies of weekly entangled states are used by Alice and Bob for distillation. By combining all these states, and applying local operation using quantum gates, pure Bell pairs can be generated \cite{hu2021long}.}  \textcolor{black}{So, a two-way classical communication is required to acknowledge both nodes about successful purification.} 
\end{itemize}

\textcolor{black}{One limitation of first generation quantum repeaters is that they require two-way classical communication for entanglement purification step, which creates a performance bottleneck as the distance of communication is increased. Thus, there is a need to move towards schemes that require only one-way communication and no classical signals are required for acknowledgement of success rate. Therefore, the \textbf{second generation} of quantum repeaters employ QEC (one-way classical communication) for the purpose of entanglement purification \cite{jiang2009quantum}, \cite{munro2010quantum}, \cite{fowler2010surface}. With this scheme, the waiting time for receiving the acknowledgement of purification is not needed, thus qubits can be used for entangling the link again without waiting \cite{munro2015inside}. Reference \cite{jiang2009quantum} uses Calderbank-Steane-Shor (CSS) encoding for significant improvements in speed of entanglement generation between nodes. But, second generation quantum repeaters using heralded entanglement generation still needs two-way signalling to reduce loss errors. Therefore, in the \textbf{third generation} of quantum repeaters, this two-way signalling is replaced by one-way signalling using certain loss-tolerant codes \cite{munro2012quantum}, \cite{fowler2010surface}, \cite{muralidharan2014ultrafast}, \cite{azuma2015all}. In \cite{fowler2010surface}, quantum information is encoded in matter qubits that are transferred to photons and are sent through the channel from the transmitter to the receiver \cite{munro2012quantum}. At the receiver, these quantum states are transferred to matter qubits before carrying out the error correction. The matter qubits are not required to serve as quantum memories as
in first and second generations. In 2014, this scheme was termed as fully fault-tolerant \cite{muralidharan2014ultrafast}. Matter qubits may be too demanding for use in quantum repeaters, and thus,
all optical quantum repeaters were developed by taking a time reversal of the DLCZ-like (discussed later in this section) quantum repeater protocol \cite{azuma2015all}.} 

\textcolor{black}{A lot of research is going on in this field with single quantum systems \cite{rozpkedek2019near}, \cite{santra2019quantum}, \cite{krutyanskiy2019light}. Significant theoretical as well as experimental progress on repeater protocols with repeater elements has taken place \cite{munro2015inside}, \cite{muralidharan2016optimal}, \cite{sangouard2011quantum}. The majority of implementations follows encoding, where information is encoded in discrete variables and protocols based on this have been existed and been improved upon since two decades \cite{munro2015inside}, \cite{muralidharan2016optimal}. But, repeaters with continuous variable \textbf{(Sec. \ref{sec: Sec. VII-D})} protocols has been a state of subject recently and the first protocol was published in 2017, 2018 \cite{dias2017quantum}, \cite{furrer2018repeaters}.}

\begin{figure}[!t]
\centering
\includegraphics[width=0.5\textwidth,height=6.5cm]{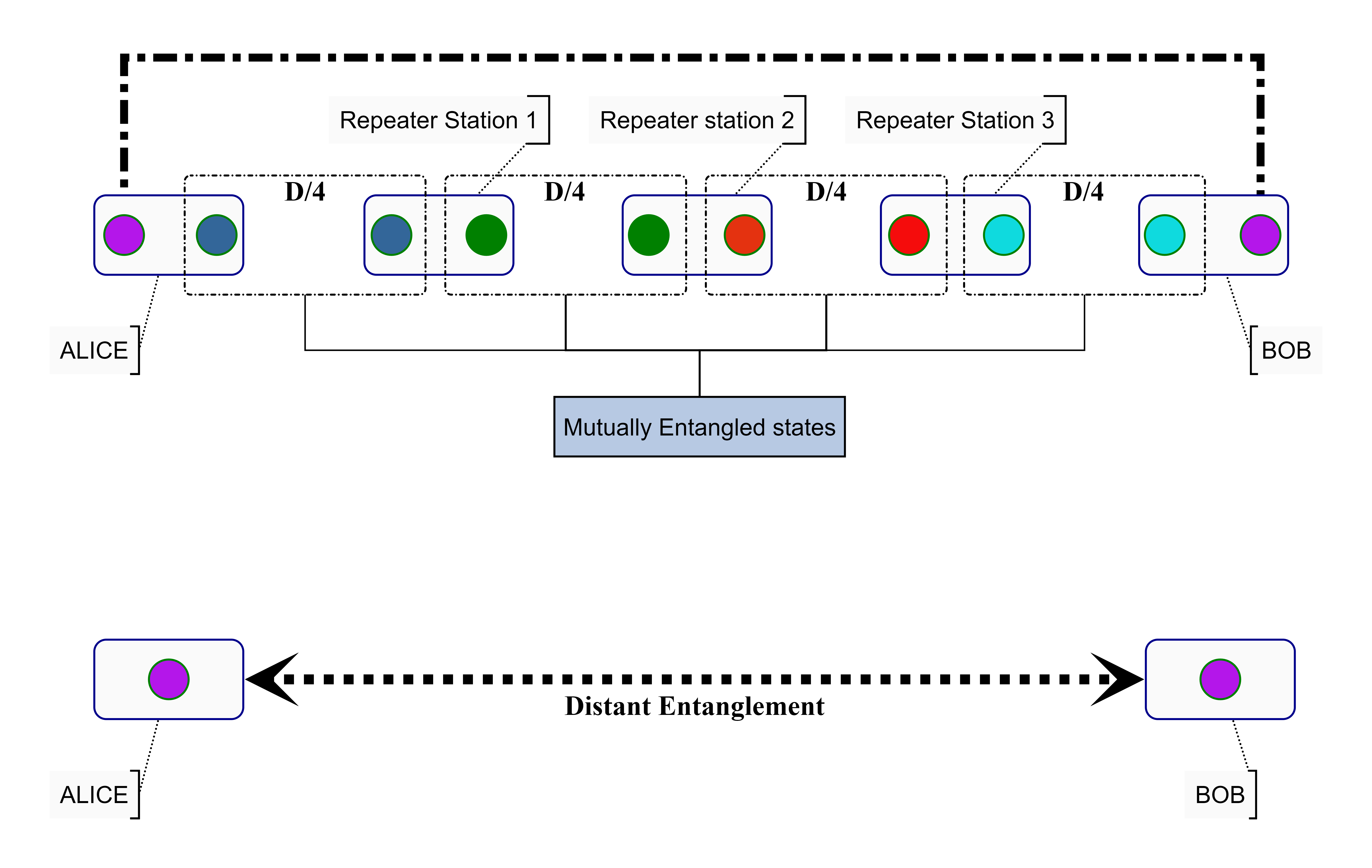}
\caption{Quantum Repeaters: Entanglement is generated between distant nodes by employing intermediary repeater stations which have qubits that are mutually entangled to each other. Thus, after performing  \textit{Entanglement swapping} operation , Alice  and  Bob  qubits  are  entangled. In this way, long distance communication would be possible which is the main requirement of \textcolor{black}{QI}.}
    \label{fig:Fig 12}
\end{figure}

There are certain conditions that are required for the design of quantum repeaters, although the losses surge exponentially with the distance of communication. Supposing, the distance of communication between Alice and Bob is \textcolor{black}{\textit{D}}. The idea is to create an entanglement between two pairs, each covering the distance of \textcolor{black}{\textit{D/2}}. Entanglement between these two pairs is created by entanglement swapping between another two pairs covering the distance \textcolor{black}{\textit{D/4}} and so on. This process is repeated unless entanglement is created between all the intermediary nodes, ultimately entangling the qubits at the far away end nodes, as shown in \textbf{Fig. \ref{fig:Fig 12}}.
The quantum repeater stations improve the photonic qubit transfer rate \cite{munro2015inside}. End nodes can also be used as quantum repeaters, thus posing similar requirements on protocols for its realization. The construction of maximally entangled states over long distances is achieved according to special protocols as described in \cite{briegel1998quantum}, \cite{bose1999proposal}. The most famous protocol is \textbf{DLCZ protocol} named after authors Duan, Lukin, Cirac, and Zoller \cite{duan2001long}. 
Particularly, adopting atomic ensembles as quantum memories and linear optical techniques, as compared to single atoms, makes it easier to maintain strong coupling between the photons and memories that prove advantageous in long-distance communication of quantum information \cite{chou2004single}. 
The idea behind the functionality of the DLCZ protocol is a photon having spontaneous Raman emission, generating a simultaneous joint spin excitation in the atomic ensemble. 
The connection between the atomic excitation in each ensemble and emitted photons forms the foundation for the generation of entanglement for each intermediary link,  done via a single-photon detection \cite{van2003atomic}.
The spin excitations can be effectively reconverted into photons through associative interference effect. This correlation between emitted photons and atomic excitations in each ensemble, forms the crux of the entanglement swapping operation. Due to these advantages offered by atomic ensembles, DLCZ protocol is experimented successfully in \cite{kuzmich2003generation}, \cite{van2003atomic}, \cite{matsukevich2004quantum}, \cite{chou2005measurement}, \cite{chou2007functional}, \cite{yuan2008experimental}. Many researchers motivated by the features of atomic ensembles and the experimental results, started improvising DLCZ protocol and came out with exciting results. Thus, knowing the basics of DLCZ protocol gives insights about how this protocol can be improved further for quantum repeater purposes. The physics behind this protocol and the entanglement creation between long distant ensembles is thoroughly discussed in \cite{sangouard2011quantum}. 
In an ensemble of three-level systems with two ground states $g_{1}$ and $g_{2}$ and excited state \textit{e}, the transition of $g_{1}$-\textit{e} is first incidented with an off-resonant laser pulse resulting in the spontaneous emission of a Raman photon on the \textit{e}-$g_{2}$ transition. This photon \textcolor{black}{which is \textcolor{black}{generated} on a condition} that energy of \textcolor{black}{\textit{$g_{2}$} is higher} than
that of \textcolor{black}{\textit{$g_{1}$}}, is termed as \textit{Stokes} photon. These photons, transmitted through optical fibers, are responsible for entanglement creation between two remote ensemble locations using DLCZ protocol.

States having minimum entanglement are termed as noisy, so after entanglement distillation, they are turned to purified states having maximum entanglement. Realizing \textcolor{black}{QI} with QKD and trusted repeater networks, requires the use of quantum satellites such as Micius \cite{yin2017satellite}, which bolster free-space quantum channels covering longer distances. The vacuum in outer space provides noiseless communication thus improving fidelity for such quantum-satellites based quantum systems. The misfortunes over longer distances, including those brought about by climatic diffraction, are significantly less. All things considered, the advancement of quantum repeaters for a quantum web requires complex quantum innovation.

\subsubsection{Workflow} 
\label{sec: Sec. IV-B-1}
The working principle of the first generation of quantum repeaters can be best understood with the help of a simplified example where the sender Alice and the destination Bob perform quantum cryptography with received entangled photons. As of now, these two end nodes are distant apart and photon losses occurring in channels deteriorated the photon rate. So, they have to keep the noise as minimised, in the channel, as possible. To achieve that, they use quantum repeater that can be expressed mathematically as 
\begin{equation}
    S_{1}\Longleftrightarrow S_{2} \big\langle b^{m}\big\rangle S_{3}\Longleftrightarrow S_{4},
\end{equation}
where “$\Longleftrightarrow$” corresponds to entanglement and “$\big\langle b^{m}\big\rangle$” corresponds to joint BSM for performing entanglement swapping discussed in what follows. 

\subsubsection{Entanglement swapping}
In this process, the projection of the state of two particles is made onto an entangled state, for obtaining an overall entanglement. It is done by performing special \textcolor{black}{BSM, which does not essentially require an immediate cooperation between the two particles, such a measurement will automatically collapse the state of the remaining two particles into an entangled state.} Entanglement swapping is nothing but the quantum teleportation of entanglement of quantum states. Multiple teleportation procedures are performed by multiple repeater stations, thus forming a quantum channel for transmission of quantum information between end nodes, that are thousands of kilometers apart physically \cite{bennett1993teleporting}, \cite{zukowski1993event}, \cite{vedral1998entanglement}.

\subsection{Quantum Channels}
\label{sec: Sec. IV-C}
The \textcolor{black}{QI} is an inherently secure \textcolor{black}{network that will be implemented along with classical internet, where end nodes will be connected via classical and quantum channels.} A continuous quantum channel is created until all the end nodes are connected \textcolor{black}{as shown in \textbf{Fig. \ref{fig:Fig 2}}.} The classical information as well as quantum information can be sent through quantum channels. As seen in teleportation of data qubit $\big|\phi_D\big\rangle$, EPR pairs that are generated, are distributed to Alice and Bob through a quantum channel. Now, this channel can be a free space channel for sending quantum information to long distances or can be fiber based channels, which can be used to send quantum information to relatively shorter distances. How efficiently an entanglement can be distributed depends upon the capacity of a channel and entanglement distribution process, which is mainly limited by the photon loss in channel that scales exponentially in a usual scenario. The communication channel capacity defines the ability of the channel to deliver the information from Alice to Bob in a faithful and recoverable way. 

The steps of quantum communication through quantum channels differ from the communication through classical channels in many ways. In quantum channel, communication of quantum information takes place in three different phases:
\begin{itemize}
    \item \textbf{Channel encoding phase}, where sender Alice encodes or prepares the quantum state to be sent to the destination Bob, so that the information encoded compensates for the noise in channel \cite{shang2014quantum}, \cite{nguyen2017towards}.
    \item Next Phase is \textbf{quantum evolution phase}, where quantum information is put on the quantum channel, which maps the information to the receiver Bob.
    \item Final phase is \textbf{decoding phase}, which includes quantum information to be decoded by destination Bob. Here received quantum state, containing the exact information, can be a superposed state or a state possessing some noise. This state is then measured by Bob taking into account the error correcting procedures.  
\end{itemize}

\subsubsection{Free Space Channels}
\label{sec: Sec. IV-C-1}
\textcolor{black}{The free-space optical channel was first utilized for QKD in \cite{hughes2002practical}, \cite{kurtsiefer2002step} by the Los Alamos National Laboratory (LANL). Some QKD experiments were demonstrated that were capable to operate in daylight conditions \cite{benton2010compact}, \cite{peloso2009daylight}. Another project sponsored by U.S. government has reportedly performed a QKD experiment with an aircraft flying at 10,000 feet to ground station using free-space channel \cite{tunick2010review}.} The Free space channel link has been demonstrated in various experiments \cite{tunick2010review}, \cite{gyongyosi2018survey}. DARPA sponsored QKD network using free space link has been demonstrated in \cite{elliott2005current}. Quantum communication has been demonstrated by the Zeilinger group between two peaks in the \textcolor{black}{Canary Islands}, separated by 144 km \cite{ursin2007entanglement}, \cite{fedrizzi2009high}. The free-space optical quantum channel has all the earmarks of being very promising for long-distance quantum communication. The information-carrying photons are quick and vigorous, and are prone to environmental loss. The advantage of using free space channel is that, most of the photon loss mechanism occurs lower than 10-30 km of atmosphere, leaving behind most of path as vacuum with negligible loss of photons and decoherence. Only dispersion effect is present once light leaves the atmosphere \cite{villoresi2008experimental}. \textcolor{black}{Thus,} free-space optical channel keeps the quantum information in photons intact and increases the system fidelity and channel capacity. \textcolor{black}{Therefore, a high information capacity photons are required for this purpose,  i.e. orbital angular momentum (OAM) modal basis of photons \cite{allen1992orbital}. QKD protocols encoding by OAM modes were verified  experimentally in the laboratory conditions \cite{mafu2013higher}, \cite{vallone2014free}. The effect of atmospheric turbulence on OAM modes are discussed in \cite{goyal2016effect}. Some research groups have developed the ways to mitigate atmospheric turbulence in free space with certain techniques such as, using adaptive optics post-processing \cite{zhao2012aberration}, \cite{willner2015optical}, robust states \cite{brunner2013robust}, and optimal encoding scheme \cite{alonso2013protecting}.}

Until the launch of \textcolor{black}{Micius satellite by China} \cite{yin2017satellite}, entanglement distribution was only possible with distances not more than \textcolor{black}{100 km} \cite{yin2012quantum}. But with this launch, it became possible to achieve entanglement distribution between two far away ground stations located on earth with distances greater than 1000 km through terrestrial free space channels \cite{aspelmeyer2003long}. Much of the system fidelity depends on the quantum channel loss, which comprises of beam diffraction, pointing error, atmospheric turbulence, absorption, beam divergence \cite{krenn2016quantum}. Thus, satellite based entanglement distribution puts more stringent requirements for link efficiencies than any classical satellite to ground communications. \textcolor{black}{Taking into account all these losses, the attenuation of the link should not be greater than 60 dB for ground-to-satellite quantum communication to be successful \cite{sharma2019analysis}. These losses leads to beam broadening which results in loss of photons limiting the photons received by receiver end.  Typically losses are in the order of 30-40 dB \cite{krenn2016quantum}. So, various techniques are heralded to achieve efficient long distance communications \cite{willner2015optical}, \cite{brunner2013robust}, \cite{alonso2013protecting}.}   


\subsubsection{Fiber Based Channels}
\label{sec: Sec. IV-C-2}
\textcolor{black}{Besides free space links, which plays a vital role in long-haul quantum communication, optical fiber links are more attractive because of their technological maturity owing to traditional internet technology. The qualitative features of both these channels are mentioned in \textbf{Table \ref{tab: IV}}. Here qubits carrying information are transmitted from Alice to Bob through optical modes in fibers. A standard single-mode fiber (SMF) consists of a core and a cladding that carries the light. Total internal reflection guides the light along the core of the fiber because of the difference in index of refraction. These waveguides are fabricated directly in silicon \cite{ballato2008silicon} or in conductors such as aluminum \cite{dhar2010fabrication}. There are different types of fibers available for propagation of quantum states; SMFs, \textcolor{black}{multi-mode} fibers (MMFs), few-mode fibers, multi-core fibers (MCFs) etc. Among all, SMFs have a smaller core that keeps the light more tightly, therefore better suited for high-precision environments of quantum experiments \cite{cozzolino2019high}.} In one such experiments, entanglement is created at a distance of 300 km through optical fiber channel \cite{inagaki2013entanglement}. 
\newcolumntype{P}[1]{>{\arraybackslash}p{#1}}
\begin{table}[ht!]
\centering
\caption{Features of Free-Space and Fiber-Based channels for Quantum Communication}
 \begin{tabular}[c]{|P{4cm}|P{4cm}| }
\hline

\textbf{Free-Space Channels} & \textbf{Fiber-Based Channels} \\

\hline \hline
It uses free space as a mode of communication. & It consists of core and cladding that carries light photons. \\
\hline
 It suffers from loss due to environmental effects such as beam  diffraction, pointing error, atmospheric turbulence, and absorption but is resistant to such losses after entering vacuum. & It suffers from absorption and coupling losses but has negligible loss due to environmental factors.\\
 \hline
Supports satellite communication where it can act as Up-link or Down-link channel. & Supports ground to ground long distance communication. \\
 \hline
 High Fidelity & Low Fidelity\\
 \hline
\end{tabular}
\label{tab: IV}
\end{table}

Fiber-based channels have an advantage that environmental factors are not significant inside fiber channels, thus preserving the coherence of photons. Instead, it suffers from polarization-preservation and optical attenuation problems that limits the distance of communication to a few hundreds of kilometres \cite{takesue2007quantum}, \cite{huang2016long}, \cite{yin2016measurement}.

\subsection{Quantum Memories}
\label{sec: Sec. IV-D}
Quantum memories are an integral part of quantum communication system, which is required by intermediary devices to store and process the qubits for sending and receiving photons with high fidelity \cite{ma2020optical}. Quantum computing includes a tool for the synchronization of the processes within a quantum computer \cite{botsinis2018quantum}. In addition, quantum memories are employed with quantum repeaters for extending the range of quantum communication \cite{briegel1998quantum}. Quantum memories were also studied in the strategic report on quantum information processing and communication, which proved to be helpful in European integrated project Qubit Applications (QAP). This had a sub-project involving seven research groups devoted to the development of quantum memories \cite{simon2010quantum}. For achieving the scalability required for large-scale long-distance quantum communication, quantum memories are employed for converting qubits between light and matter. They also control the storage of qubits and retrieve them with stationary matter systems \cite{simon2010quantum}, \cite{duan2001long}. Quantum memories are also required for timely operations for perfect timing and synchronization \cite{Kimble2008}, \cite{duan2010colloquium}, \cite{sangouard2011quantum}, \cite{simon2010quantum}. As discussed earlier, communication between distant nodes storing matter qubits, engages quantum memories for the transfer of quantum information between these nodes with entanglement distribution \cite{chou2007functional}, in free space or fiber based channels \cite{stenholm1998teleportation}. For this, atomic ensembles were proved to be best contender for quantum memory, with good amount of experimental progress \cite{julsgaard2001experimental}, \cite{chou2005measurement}. Quantum memories having cold neutral atoms are discussed in \cite{ding2018raman}, \cite{choi2008mapping}, having doped crystals in \cite{usmani2012heralded}, enabling the storage and retrieval functionality of single-photon entanglement extending to high-dimensional entanglement \cite{zhou2015quantum} and continuous-variable entanglement \cite{jensen2011quantum}. As a result, transfer efficiency of all these implementations lies between 15\% and 25\% \cite{yan2017establishing}. High fidelity entanglement transfer is a major challenge for network scalability inspite of various implementations of efficient quantum memories for polarization qubits \cite{vernaz2018highly}, \cite{wang2019efficient}.

\subsection{Quantum Key Distribution}
\label{sec: Sec. IV-E}
Every field in today's modern era, whether its military applications, financial institutions, private multinational companies or government agencies etc, rely upon a full-fledged internet to accomplish the day to day tasks and shares several Gigabytes of data over the internet. The security of this data is fundamental, as any eavesdropping attack on private/public channels, will be a serious attack on the privacy as shown in \textbf{Fig. \ref{fig:Fig 1}}. The \textcolor{black}{security of systems using traditional cryptographic protocols is based upon the complexity in solving mathematical problems} such as factorization of large integers or discrete logarithmic problem, and believing that the hackers computing performance is inadequate. \textcolor{black}{Cryptography is classified into symmetric, asymmetric and hybrid cryptography where some type of keys are used for encryption and decryption of messages. If the same key is used for encryption and decryption of messages, such systems are known as \textit{private key cryptosystems} \cite{rao1986private}. On the other hand, \textit{public key cryptosystems} use public key for encryption and private/secret key for decryption \cite{okamoto2000quantum}. The idea of public key cryptosystem was put forth by Diffie-Hellman in 1976 \cite{diffie1976new}. The first implementation was done by Rivest, Shamir, and Adleman (RSA) in 1978 \cite{rivest1978method}. Since, public key cryptosystems are slow, hybrid schemes are employed that combines private key and public key methods offering considerable speed advantages \cite{dent2004hybrid}. In public key cryptosystems, without the knowledge of the secret key a message cannot be decoded, thus enhancing the security of the systems. Unfortunately, private key cryptosystems suffers from \textit{key distribution problem},  i.e. distributing the keys in a secure manner.} QKD provides excellent privacy against hacks, and provides solution to key distribution problems using quantum cryptography \cite{bennett1984quantum} with the following objectives:
\begin{itemize}
\item First, sender Alice and Bob shares a secret/private key with each other, without \textit{Eve} having no knowledge of it.
\item Next, this key is distributed from Alice to Bob through an inherently secure quantum channel. The security of this key depends upon the fact that any attempt of hacking or spying by \textit{Eve} will disturb the state of the key qubits, thus exposing his/her attempt. Hence, it is impossible for \textit{Eve} to copy the information without being detected.
\item Alice and Bob must be capable of \textit{authenticating} the message, in order to prevent it from \textit{Eve} taking the position of either Alice or Bob. \textcolor{black}{This task can be accomplished by quantum authentication protocols such as \cite{zeng2000quantum}, \cite{barnum1999quantum}, \cite{duvsek1999quantum}.} If such level of authentication can be achieved, QKD can provide provable security against unlimited attacks, that is nearly impossible with classical key exchange cryptographic protocols.
\end{itemize}

\textcolor{black}{The first experiment to demonstrate QKD was performed in 1989, covering distance of 32 cm \cite{bennett1989experimental}. This distance has to be increased practically for achieving QKD over long distance using optical fiber channels or free space channels. But, with increased distance, fiber and atmospheric losses increases, resulting in decrease of photon count and key rate significantly due to decoherence effect \cite{cao2019kaas}, \cite{cao2020multi}. No-cloning theorem \cite{wootters1982single} also restricts sending several copies of information, thus limiting the practically achieved distance of QKD to few hundreds of km \cite{yin2016measurement}. Therefore, researchers are focused on exploring satellite to ground QKD so that two remote nodes or network of nodes achieves QKD for long distance without significant absorption and turbulence \cite{rarity2002ground}. Following this, in 2004, an experiment was conducted in free space that reported a successful survival of entanglement of photons with noisy atmosphere of 13 km from ground \cite{peng2005experimental}. This distance was further improved from 100 km to upto 500 km with various satellite based successful QKD experiments \cite{ursin2007entanglement}, \cite{yin2012quantum}, \cite{wang2013direct}, \cite{nauerth2013air}. With the launch of Micius satellite, breakthrough has been achieved with QKD covering much greater distance of 1200 km \cite{yin2017satellite}. Using same satellites, a secure key has been generated between two ground stations located in China and Europe that are 7600 km apart \cite{liao2018satellite}. In addition, already existing fiber links of 2000 km,  i.e. that between Beĳing to Shanghai, are integrated with free space QKD links using Micius. A quantum communication distance of 4600 km was reported between ground stations \cite{chen2021integrated}. These experiments provide significant breakthrough in realizing global space-based QI.}

\subsubsection{Key Protocols}
\label{sec: Sec. IV-E-1}
For providing security and integrity of data, QKD systems operate in two stages. 
In the \textbf{first stage}, a random bit sequence, known as “key”, is generated, which can be a result of certain binary operation performed on the data to be transmitted. In the \textbf{second stage}, the generated quantum “key” is sent to destination in a secure manner over classical channel where it can be decoded using the same private “key”. A cryptographic protocol used in classical systems is OTP, which is provably secure in quantum technology \cite{shannon1949communication}. No algorithm will be able to acquire the knowledge of the “key”, as it is generated quantum randomly. Some of these protocols are discussed below.
\begin{itemize}
    \item \textbf{OTP}: OTP is one of the oldest classical cryptographic techniques that dates back to 1882, it was invented by Frank Miller \cite{bellovin2011frank}. It is a simple and highly effective private key cryptosystem. Security of the OTP cannot be compromised without the knowledge of a \textit{pre-shared} key, which is known only to Alice and Bob (private key). If the information is intercepted by \textit{Eve}, then Alice and Bob will get to know about the spying, as a result the information will be declared \textit{invalid}. The data is encrypted by a binary operation on \textit{data} and \textit{key} bits. The result of this operation is sent to Bob, who then decodes the information using the same binary operation on the received data bits, thus retrieving the original information with the help of a pre-shared private key having the following features:
    \begin{enumerate}
        \item The key generated must be used only once. It cannot be reused and should be in its original form. Not even one part of this key must be reused as it will compromise the security of the system.
        \item The length of the key must be at least as long as the data to be sent, so that even if \textit{Eve} attacks the channel she does not succeed.
        \item The key must be shared to Alice and Bob in a secret manner, so that only Alice and Bob must be in a possession of this key.
    \end{enumerate}
 Since the OTP requires the length of the key to be at least as long as the amount of data to be transmitted, it becomes impractical to be used in several cases. Hence, \textcolor{black}{there is} a need to explore and exploit the property of entanglement, to generate a quantum vernam cipher \cite{leung2002quantum} and other protocols for enhancing the security \cite{murray2020implementing}. 
 
 
\item \textbf{BB84 Protocol}: This protocol is one of the popular methods used for quantum cryptography. It was formulated by two physicists, Charles H. Bennett and Gilles Brassard, at IBM in 1984 \cite{bennett1984quantum}. Though original idea was proposed by Wiesner in 1970's, but the research could not be published in the same year as it was published a decade later in 1983 \cite{wiesner1983conjugate}. In this protocol, back to back photons carrying qubits are sent to Bob, who tries to perform a measurement on the arrived photons carrying qubits. Some of the photons are lost due to decoherence and never arrive while some of those are not detected. Out of all the detected photons, some of the bits are utilized to form the \textit{key}, the rest are utilized to expose the \textit{Eve} and gather information about spying. Alice encodes the information with her own \textit{basis}, about which only Bob is familiar with. If this information is decoded using a different basis, other than the Alice's one, then the measurement will \textit{disturb} the state of the photons carrying the qubits. In this context, \textit{Eve} measuring the information with different basis \textit{alters} the state of qubits, thus revealing its presence, resulting in Bob to discard those bits.
\begin{figure}[!t]
\centering
\includegraphics[width=0.5\textwidth,height=6.5cm]{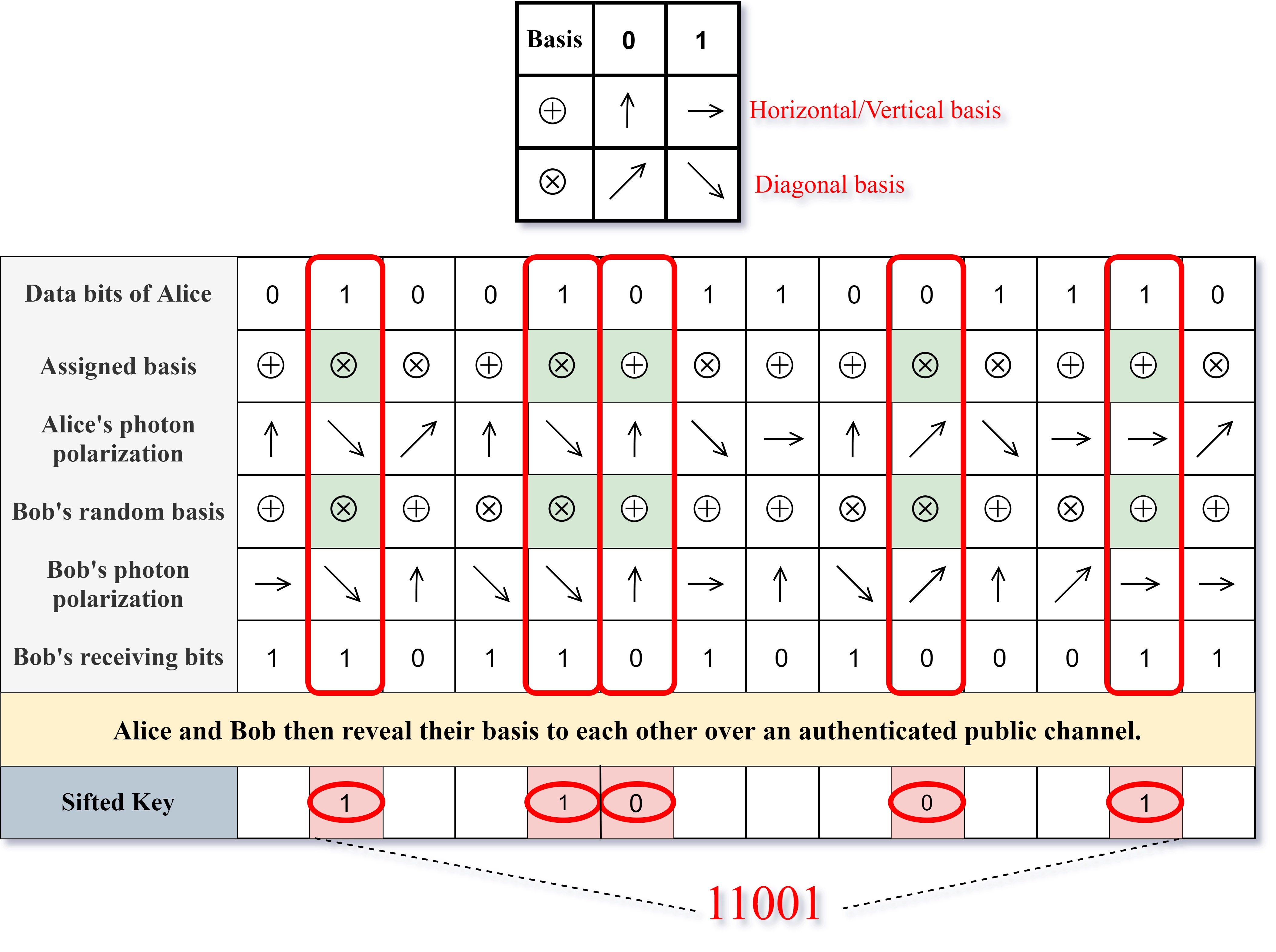}
\caption{BB84 protocol.}
    \label{fig:Fig 13}
\end{figure}
BB84 protocol is based on the communication of a series of single qubits sent from Alice to Bob, that are stored in photons represented by their polarization states. Usually, the transmission between Alice and Bob takes place through quantum channels but fiber-optic channels can also be used. Alice can use the \textit{vertical/horizontal} or \textit{diagonal} basis of polarization for encoding the information,  i.e. $\big|\uparrow\big\rangle$  for vertical basis, $\big|\rightarrow\big\rangle$ for horizontal basis, $\big|\nearrow\big\rangle$ and $\big|\searrow\big\rangle$ for diagonal basis. Let the states $\big|\uparrow\big\rangle = \big|0\big\rangle$ and $\big|\rightarrow\big\rangle = \big|1\big\rangle$ be represented by $\oplus$. The states $\big|\nearrow\big\rangle = \big|0\big\rangle$ and $\big|\searrow\big\rangle = \big|1\big\rangle$ are \textcolor{black}{represented by $\otimes$.} The protocol runs as follows:
    \begin{enumerate}
        \item The quantum states are encoded using the data bits of Alice with the assigned basis (e.g. $\oplus$ \textcolor{black}{and $\otimes$)} and random classical bits (0 or 1). Alice sends the encoded quantum states to Bob through any channel that is usually vulnerable to attack by \textit{Eve} for spying. The data encoded is in the form $\big|\uparrow\big\rangle$, $\big|\rightarrow\big\rangle$, $\big|\nearrow\big\rangle$, $\big|\searrow\big\rangle$.
        \item Bob then uses a random basis for measurement of qubits received with basis $\oplus$ and $\otimes$ as shown in \textbf{Fig. \ref{fig:Fig 13}}. This is the last quantum operation carried out in this protocol. 
        \item Bob then acknowledges to Alice for the qubits that he has received via a public channel. 
        \item Alice and Bob discloses the basis used to encode the data qubits via a public channel. When both choose the same basis, the corresponding measured bits are kept and this data will form the \textit{key}. If they choose a different basis then the bits will be \textit{rejected} and this process is well known as \textit{sifting}.
        \item Next, the presence of \textit{Eve} is checked by revealing a random number of bits encoded by Alice. If the detected bits by Bob are the same, then there is no interference by \textit{Eve} and the remaining error free bits are termed as \textit{encryption keys}.
        \end{enumerate}
    
In 2007, results from polling stations in Swiss parliamentary elections, in the canton of Geneva were transferred, using this protocol, to Bern over hundreds of kilometres and nobody could hack the system till date. In this way, a secure communication is possible in any distributed network exploiting the realms of quantum mechanics. It is used in satellite based communications theoretically and experimentally as shown in \cite{fedrizzi2009high}, \cite{aspelmeyer2003long}, \cite{villoresi2008experimental} and \cite{wang2013direct}. \textcolor{black}{This reach can possibly be extended using \textit{relays}, that may or may not be trusted \cite{yin2020entanglement}.} In future QI, QKD will play a very important role in making the internet hack-proof as it is the main requirement in today's digital era. But every solution comes with some shortcomings and BB84 protocol is no different. The biggest disadvantage is its limited distance upto which it can work efficiently. It can cover only up to hundreds of km through fiber-based or free space atmospheric channels.
\item \textbf{Ekert Protocol}: This protocol which takes into consideration the entanglement of photons carrying qubits, was proposed in 1991 by Artur Ekert \cite{ekert1991quantum}. The first implementation of this protocol was achieved by Anton Zeilinger and his group in 1999 over a short distance of 360 m \cite{bouwmeester1999observation}. By the same team, \textcolor{black}{the first money transfer via quantum cryptography was demonstrated in 2004 \cite{beth2005cryptanalysis}.} This protocol runs in 6 steps as shown in \textbf{Fig. \ref{fig:Fig 14}}:
\begin{enumerate}
    \item First, an authenticated channel must be set up, which can be classical or quantum in nature, between Alice and \textcolor{black}{Bob} and they have to make sure there is no chance for \textit{Eve} to take position of any one of them. This channel is open to interceptions. 
    \item  Next, an EPR pair of entangled photons is sent to Alice and Bob, whose measurement results in values of 0 or 1. Alice and Bob filter those bits that shows clear correlations (important bits) from those without clear correlations (unessential bits). For this, need of classical channel is a necessity. 
    \item Then, privacy amplification procedure \cite{deutsch1996quantum} is employed for correlation of bits owing to measurement errors, that are unavoidable in practice, are correct them.
    \item The network security depends upon whether the states are maximally entangled or not, otherwise decoherence becomes an obstacle in the communication. \textit{Eve's} hacking attempt can be detected with \textcolor{black}{Bell's inequality} \cite{aspect1999bell}. If this inequality is not violated, then it can be believed that there is an eavesdropping attack, or the information communicated has some technical issues. As a result, any successful attack by \textit{Eve} can be detected immediately and the system functionality is checked.
    \item The quantum key, taking into consideration the relevant bits, is used for a symmetrical procedure such as OTP and sent over the regular internet.
    \item After that, Bob receives the data that are encrypted in a ciphertext and deciphers it with the self generated private key. This completes the Ekert protocol resulting in the generation of a secure key. 
    \end{enumerate}
\end{itemize}
\begin{figure}[!t]
\centering
\includegraphics[width=0.49\textwidth,height=7.5cm]{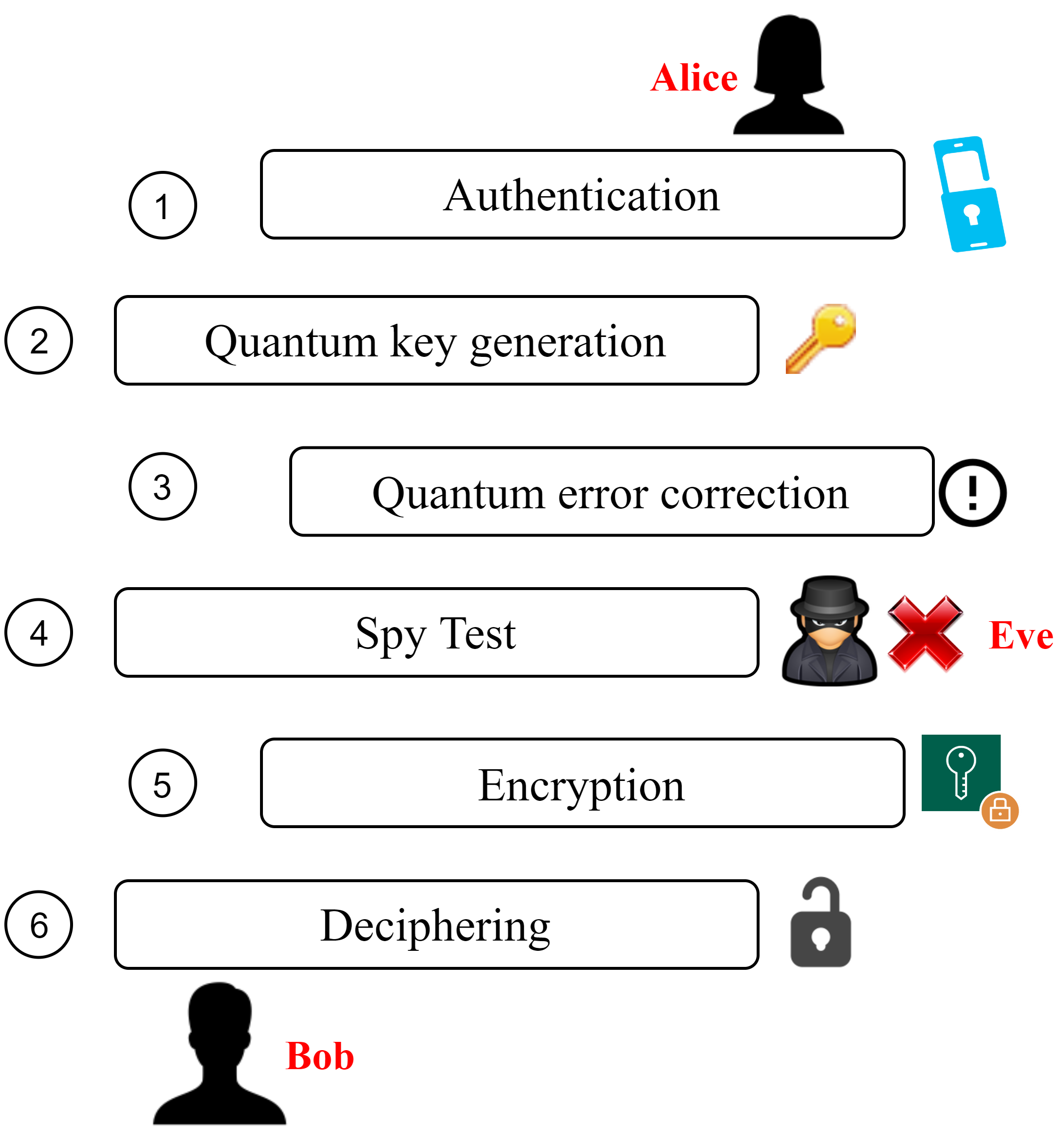}
\caption{An example of Ekert protocol: This entanglement-based QKD protocol runs in 6 steps.}
    \label{fig:Fig 14}
\end{figure}

\subsubsection{Security of QKD}
\label{sec: Sec. IV-E-2}
\textcolor{black}{Quantum key distribution} relies upon protocols and the specified steps of operations to be performed between Alice and Bob. It is generally a two-step process. In the first step, Alice and Bob takes some correlated data, e.g. \textit{X} or \textit{Y}, that can be quantum or classical in nature. Then, in the next step, Alice and Bob exchange some information through classical or quantum channel, resulting in an output \textit{X'} and \textit{Y'}. This output is generally a pair of secret keys that are only shared between Alice and Bob. If uncorrelated data inputs are used, then a protocol might not be able to produce the desired output,  i.e. \textit{secret keys}. For any protocol to be secure, depending on its input, it should be able to judge the situation and abort the generation of keys in case of tampered inputs. This explains that a QKD protocol must be \textit{robust} enough so that it does not compromise the security of the system.

The protocols discussed in \textbf{Sec. \ref{sec: Sec. IV-E}-1} are provably secure. The proof remains in the observation that after \textit{information reconciliation} \cite{brassard1993secret} and \textit{privacy amplification} \cite{lomonaco1999quick} are performed, the key rates obtained turns out to be coinciding with the achievable qubit transmission rate over noisy communication channels \cite{mehic2020error}. So, it can be concluded that the OTP, BB84, Ekert protocol, and EPR protocols \cite{lomonaco1999quick} are secure as long as \textit{Eve} attacks on the transmission of one qubit at a time. If \textit{Eve} performs simultaneous attacks where she obstructs and stores large blocks of the qubits to be communicated, then it will compromise the security of the system. But then again, QEC codes plays an important role in maintaining the security of these systems \cite{knill1997theory}.

There can be a case when Alice and Bob share imperfect keys and they both have upper bound on \textit{Eve}. Therefore, keys cannot be correlated, thereby compromising the security which results in \textit{Eve} intercepting the channel and gaining access to private data. To enhance the security, \textit{information reconciliation} followed by \textit{privacy amplification} \cite{bennett1995generalized}, \cite{bennett1988privacy} steps increases the correlation of the private keys and minimizing the \textit{Eve's} chances of accessing the information, achieving desired level of security. 

The quantum realms inherits the features of secure communication such as no-cloning theorem. Hence, QKD protocols governed by the \textcolor{black}{laws of quantum mechanics} while sharing the keys between Alice and \textcolor{black}{Bob} on private or public channel, are inherently hack-proof. An \textit{eavesdropper} can attack any communication between Alice and Bob in different manner. She can attack either quantum or classical channel and compromise the security. 
Another case can be taken where \textit{Eve} likely to gain the information, tries to measure the intercepted photons before sending the photons Bob. \textit{Eve} is still unaware of the basis measure each bit of photons with wrong basis having 50\% probability. Hence, this wrong measurement changes the state of the information carrying photons, before Bob receives it. This measurement will affect the ability of \textcolor{black}{Bob} to measure the bits correctly because the state of the photons has already been changed by \textit{Eve}. Even if Bob uses the basis shared by Alice to measure the photons, he will still measure the photons correctly with 50\% probability. Therefore, every time a photon qubit is measured by \textit{Eve} before sending it to Bob, Bob will be left with 75\% probability that he will measure the bits correctly and 25\% chances that he will measure the bits wrongly. Therefore, it can be seen that a considerable amount of error is introduced by \textit{Eve} by attacking the quantum channel. Hence, there is 25\% probability that \textit{Eve's} extracted key is incorrect, and eavesdropping can also be exposed by Alice and Bob. Therefore, a private key will be generated or an eavesdropping will be detected. So, to overcome this obstruction, \textit{Eve} has the option of copying the qubit, and sending the original bit to \textcolor{black}{Bob} while keeping the copy to herself. 
\begin{table}[ht!]
\caption{End node complexity perspectives}
\centering
\label{tab: Table V}
\begin{tabular}{@{}lc@{}}
\toprule \toprule
\textbf{End-node type}          & \textbf{End-node functionality} \\ \midrule
Single qubit operation & QKD                    \\\midrule
Single qubit + memory                & \begin{tabular}[l]{@{}l@{}}Access to connected \\ quantum computers\end{tabular} \\\midrule
Multiple qubits (today's scale)      & \begin{tabular}[l]{@{}l@{}}Emerging distributed\\ computing\end{tabular}         \\\midrule
Full-scale (future) quantum computer & \begin{tabular}[l]{@{}l@{}}Distributed quantum\\ computing\end{tabular}          \\ \bottomrule\bottomrule
\end{tabular}%
\end{table}
The qubit which is kept by \textit{Eve} can be measured later on when Alice and Bob discloses their basis of measurements. But this idea is not practical in quantum realms because of the property of no-cloning. \textcolor{black}{Hence, any attempt to copy the qubit will be detected keeping the security of QKD protocols intact \cite{nurhadi2018quantum}.} But in such protocols, it is possible that \textit{Eve} pretends to be, Bob to Alice, and pretends to be, Alice to Bob. Therefore, proper \textit{authentication} protocols are required such as \cite{zeng2000quantum}, \cite{barnum1999quantum}, \cite{duvsek1999quantum}, \cite{li2004quantum}, \cite{jensen2000quantum}, \cite{ljunggren2000authority} to guard against such an attack. Also, QEC codes \cite{knill1997theory} are there to mitigate various types of attacks by \textit{Eve}, making the communication system secure and sound. The amount of information \textit{Eve} can gain is reduced significantly using these codes and also noise is compensated in channels. 

\subsection{End Nodes} 
\label{sec: Sec. IV-F}
End nodes are the most basic and the most important requirement of \textcolor{black}{quantum networks}. All the components of practical QI need \textit{end nodes} for long-haul communication to take place. These nodes can be future quantum computers, quantum memories, quantum repeaters, and classical computers working in synergy with future quantum computers. Fidelity can be improved with the help of multiple qubits employed at the end nodes making it a small scale quantum computer for communication purposes. These nodes have different set of functionalities ranging from QKD to distributed quantum computing. An end node type and its corresponding functionality is presented in \textbf{Table \ref{tab: Table V}}. Quantum computers were first introduced way back in 1980's by Feynman in \cite{feynman1982simulating}, there it was proposed that it is possible for one quantum device to work in synergy with other quantum devices that are more efficient than any classical computer. A quantum turing machine was suggested by Paul
Benioff back in 1982 \cite{benioff1982quantum}. David Deutsch formulated the first quantum algorithm based on quantum Turing machines in \cite{deutsch1985quantum}, \cite{deutsch1992rapid}. With all these advancements, the first implementation of real quantum computer was done in 1993 \cite{lloyd1993potentially}. Also in 1994, Peter Shor in his pioneer research developed a prime factorization algorithm that can factor composite numbers or take discrete logarithms in time polynomial \cite{shor1999polynomial}. 
\textcolor{black}{Therefore, with such a huge computation power, various encryption and security protocols such as RSA \cite{rivest1978method} and Diffie-Hellman key exchange \cite{steiner1996diffie} are at risk. This sets stringent requirements of ultra secure internet technology.}

For the realization of QI to become a reality, the end nodes must have a robust storage of quantum states, until the maximal entanglement is established between the end nodes \cite{wehner2018quantum}. For this, entanglement generators must be employed in physical layer of QI functionalities \cite{cacciapuoti2020entanglement}. 
Quantum repeater stations along with quantum memories are also realized at the end nodes for achieving long haul quantum communications \cite{simon2010quantum}. If all these robust end node types are intermingled efficaciously, the peak stage of the \textcolor{black}{QI} may become a reality soon. 

\section{POSSIBLE APPLICATIONS}
\label{sec: Sec. V}
Quantum internet is a diverse field with a possibility of a whole bunch of applications \cite{cacciapuoti2020entanglement}. As this technology is still in its infancy, it is not possible to predict the exact area of applications that it can offer. But, as this technology is developing, many applications are already predicted and proved with relevant literature. Some of them are discussed below like secure communication, blind computing, clock synchronization, longer baseline of telescopes, and internet of things. 

\subsection{Secure Communication}
\label{sec: Sec. V-A}
Secure communication is the first and foremost requirement in modern everyday life. There are numerous applications which depends on transfer of data from one node to another using internet like online banking, stock exchange trading, online shopping, military applications, national security, IoT, etc. QI with various QKD protocols will maintain the security and privacy of the information exchanged over the internet. \textcolor{black}{Traditional classical communication in today's world uses various encryption protocols, such as Diffie-Hellman key exchange \cite{steiner1996diffie} and RSA protocols \cite{sedlak1987rsa}, in e-commerce websites.} \textcolor{black}{These protocols can be cracked with future full-scale quantum computers, putting all the security systems at a risk.} But, with QI and security protocols such as, QKD \cite{brassard1993secret}, Quantum Secret Sharing (QSS) \cite{hillery1999quantum}, Quantum Secure Direct Communication (QSDC) \cite{long2002theoretically}, quantum teleportation \cite{bennett1993teleporting}, and quantum dense coding \cite{bennett1992communication}, any attempt to hack the system will be detected on-site preserving the privacy of systems.
QKD has been researched widely, and successfully applied to a distance covering from few hundreds of kilometers \cite{korzh2015provably}, \cite{chen2020sending}, \cite{fang2019surpassing} to thousands of km with the launch of Micius satellite by \textcolor{black}{scientists from China \cite{yin2017satellite}}. 

QKD can be completed non-deterministic and deterministic. Non-deterministic QKD can be observed in the BB84 protocol \cite{bennett1984quantum} and in BBM92 protocol \cite{bennett1992quantum}. 
In case of QSDC, some secret information is sent securely through a quantum channel directly, without any initial generation of keys \cite{long2002theoretically}, \cite{long2007quantum}, \cite{zhu2014cryptanalysis}. Other protocols based on QSDC are quantum signature \cite{yoon2014quantum}, quantum dialog \cite{gao2010two}, \cite{zheng2014quantum} and quantum direct secret sharing \cite{zhang2005multiparty}. The QSDC scheme is also supported by a block transmission technique formulated in \cite{long2002theoretically}. High-dimensional entanglement based QSDC protocol is formulated in \cite{wang2005quantum}, \cite{jin2011quantum}, \cite{meslouhi2013quantum}, multipartite entanglement based protocols are discussed in \cite{wang2005multi}, \cite{sun2012quantum} and hyper-entanglement based protocol is developed in \cite{tie2011high}, that explains the security of QSDC based quantum communication systems. 

As IoT plays an important role in our lives due to wide area of applications \cite{suarez2016home}, \cite{hernandez2018plug}, thus protecting data involved in wide IoT applications is most important aspect which can be possible with inherently secure QI. 
The security that relies on the actual encryption primitives is risked being destroyed by the \textcolor{black}{upcoming quantum computers that are already in development.} \textcolor{black}{Quantum attacks \cite{jain2016attacks} that includes \textit{individual, collective or general} attacks using future quantum computation capabilities \cite{xu2020secure}, affect public-key cryptosystems \cite{kerry2013digital}.} All these cryptosystems based upon the prime factorization of integers, the elliptic-curve problem of discrete logarithm or the problem of discrete logarithm \cite{shor1999polynomial}, are vulnerable to quantum attacks by sufficiently powerful quantum computers. Also, Grover’s algorithm \cite{grover1996fast} can be used for attacks on symmetric ciphers by roughly a quadratic factor \cite{mosca2000private}. \textcolor{black}{Security of IoT will benefit from quantum cryptography as it works on the principles of quantum theorems.} It can be applied both in wireless and optical communication, both of which are requirements in security and integrity of IoT systems. Post-quantum IoT cryptosystems has been surveyed in \cite{fernandez2019pre} and it can improve the safety of many related fields, that rely heavily on resource-constrained and battery-dependent IoT devices. \textcolor{black}{Post-quantum cryptography refers to cryptographic protocols that are resistant to quantum attacks \cite{jain2016attacks}. It can impact various applications of IoT, that requires critical security, such as defense and public safety \cite{fraga2016review}, industrial IoT railways \cite{fraga2017towards} or smart healthcare \cite{islam2015internet}.} Therefore, the transition from classical primitives to quantum primitives is essential to ensure the overall security of data on the way.

\subsection{Blind Computing}
\label{sec: Sec. V-B}
Blind Quantum Computation (BQC) is another application of \textcolor{black}{QI} that is related to the security of a user. In this way his privacy is not leaked to a quantum computer, which is involved in computation \cite{arrighi2006blind}. The technology to build a quantum computer is moving at a fast pace, but only a few centers in the world may use it because of its superiority and cost. Just like today's supercomputer rental system, users may receive limited access rights. Supposing a user, Alice, wants to have limited access to a mainframe quantum computer, and shares her sensitive information, for computation. The result of this computation is sent back to Alice through the classical channel. In this case, if Alice want the result of computation without revealing her sensitive information, she has to use a BQC protocol.  It allows any user to interact with a mainframe quantum computer without revealing her sensitive information to any person (say Bob) having access to mainframe quantum computer \cite{morimae2013blind}, \cite{abadi1989hiding}. Lately, e-payments, in almost every field, has become the mainstream because of e-commerce websites. However, their security is compromised by emergence of quantum computers in market. Recently, based on BQC, e-payment security evaluation system is employed for evaluating security of network transactions, that guarantees security \cite{cai2020implementation}.

BQC was originally proposed by David Chaum in 1983 \cite{chaum1983blind} and was first formulated by Childs in 2001 \cite{childs2001secure}, where encryption was secured with OTP using circuit model such that server with full-fleged quantum computer keeps the information private and learns nothing about it. The protocol proposed by Arrighi and Salvail in 2006 \cite{arrighi2006blind}, requires Alice to prepare and measure the multi-qubit entangled state. In addition, it is sensitive to cheating, which means that, if full-fledged quantum computer does not mind getting caught, he can obtain information. In such a protocol, it was required a constant-sized quantum computer having access to a quantum memory for Alice \cite{yao2003interactive}. A protocol using one-way model \cite{raussendorf2001one} was formulated in 2009 by Broadbent, Fitzsimons and Kashefi in \cite{broadbent2009universal}, where Alice only needs a classical computer and a small scale quantum device that will use randomly rotated single-qubit states for its computation. Hence, this protocol is unconditionally secure and Alice’s input and  output are kept secret. The security of this protocol is experimentally demonstrated in an optical system \cite{barz2012demonstration}. This research has inspired many other researchers to implement more robust blind quantum computation protocols. Hence, \textcolor{black}{QI} in future will deploy costly and powerful quantum computers on the cloud, majorly owned by big tech companies. However, in addition to providing security and privacy to the user data, BQC would also check whether the computation is executed by a quantum processor, and is not a fraud.

\subsection{Clock Synchronization}
\label{sec: Sec. V-C}
An important issue in many practical applications of modern technologies is accurate timekeeping and synchronization of clocks all over the world \cite{corbett2013spanner}. Due to accurate synchronization of clocks, all the scientific applications dependant upon time synchronization, shows minimal errors. These applications include quantum positioning system \cite{duan2020survey}, synchronization of electric power generators feeding into national power grids, synchronous data transfers and financial transactions, long baseline of telescopes and distributed quantum computation. 
\textcolor{black}{But, these navigation systems, such as GPS, suffers from security threat known as spoofing \cite{humphreys2008assessing}, \cite{warner2003gps}. Therefore, a scheme of distributing high precision time data through quantum Micius satellite securely is reported in \cite{dai2020towards}. It is based on two-way QKD using free-space channel.}

As high-phase coherent lasers are developing, optical atomic clocks, operated by several atoms, have reached the Standard Quantum Limit (SQL) with considerable stability. SQL is set by the number of available atoms and interrogation time \cite{bloom2014optical}, \cite{hinkley2013atomic}, \cite{nicholson2012comparison}. Beyond SQL, significant improvements can be achieved in clock performance by preparing atoms in a quantum-related state \cite{komar2014quantum}. For clock synchronization, entanglement techniques are required which connects distant quantum nodes with clocks to be connected in a quantum network \cite{cirac1997quantum}, \cite{Kimble2008}, \cite{perseguers2013distribution}. A distributed network of quantum confined clocks, separated by long distances, can be treated as a ultimate clock, where all members in different countries can merge their resources together and coherently achieve higher clock stability. In this way they can distribute this international time scale to everyone in real time. The distributed architecture permits each member of the network to profit from the stability of the local clock signal without losing sovereignty or negotiating to security.

Quantum algorithms are required for addressing the problem of Quantum Clock Synchronization (QCS). It is required to minimise the difference of time,  i.e. $\delta$, between two spatially separated clocks with least communication resources. The $\delta$ can be calculated accurately and is dependent upon the stability of clock frequency and the uncertainty of the delivery time of messages sent between the two clocks separated by long distance. Presently, protocols are being developed to measure the $\delta$ with greater accuracies. \textcolor{black}{Some protocols shows accuracies better than 100 ns for clocks separated by more than 8000 km \cite{chuang2000quantum}. With advancements in these protocols such as Ticking Qubit Handshake (TQH), as discussed in \cite{chuang2000quantum}, this accuracy can be increased to 100 ps.} It allows clock synchronization that is not dependent upon uncertainties in delivery time of message between the clocks.

\subsection {Longer Baseline of Telescope} 
\label{sec: Sec. V-D}
In modern world, astronomy totally depends on very long baseline interferometry between telescope having greater resolving power than a single telescope. Infrared and optical interferometer arrays used in today’s world \cite{monnier2003optical}, \cite{lawson2000principles} have photons incoming at different telescopes, that must be substantially gathered together for measuring the interference. Due to fluctuations in phase and photon loss in transmission, the baseline can only be limited to a few hundreds, at most. Sufficient sensitivity and improved resolution of telescopes will have many impact on various scientific applications based on space such as; measuring star distance or imaging of planets outside the solar system. 
Quantum communication using quantum repeaters can reliably send quantum states of qubits over noisy communication channels with \cite{briegel1998quantum}, that allows the teleportation of quantum states over long distances with least amount of error. Using this technology, the baseline of telescope arrays can be extended as compared to conventional telescopes arrays. If a single pair of telescopes is used, that has a fixed baseline, they would not reconstruct the original distribution of source brightness effectively. Instead of this, if an array of telescopes with different baselines is used, it will acquire much more information than usual \cite{zernike1938concept}. Thus, quantum communication can be applied to extend the baseline of telescopes to exponentially increase the amount of information gathered using traditional schemes.

\section{OPEN DIFFICULTIES AND CHALLENGES}
\label{sec: Sec. VI}
Here, we present the open difficulties and challenges that are faced in the process of realization of the \textcolor{black}{quantum networks}. We start with the problem of decoherence followed by the effects of decoherence in quantum teleportation, entanglement, design complexity and the issue of collapse of the quantum states when measured.      

\subsection{Decoherence}
\label{sec: Sec. VI-A}
Quantum communication takes place when wave functions of quantum states travel from Alice to Bob through free space or fiber based channels \cite{vallone2015experimental}. In physics, two waves are said to be \textit{coherent}, if they maintain fixed phase relationship to each other or changes, according to the laws of physics, with time. Due to environmental effects and fragility of photons carrying the qubits, the interaction \textcolor{black}{of the quantum states} with the environment causes loss of photons, and hence the phase relationships are lost. This phenomenon of loss of quantum information in the environment with time is known as \textit{decoherence}. Due to decoherence, quantum states lose their properties and enter the world of classical physics.
For preserving the quantum state of photons carrying the information, they must be perfectly-isolated from the environment. However, this isolation is not practical and is hard to achieve considering the current quantum technologies. Decoherence is irreversible and it affects the whole communication process including quantum teleportation process or entanglement generation and distribution process. Reliability of a teleportation system is measured with quantum fidelity which is considered as the fundamental figure of merit, and larger the imperfections introduced by decoherence, the lower is the fidelity. Thus, decoherence degrades the fidelity of teleportation \cite{cacciapuoti2020entanglement}, and due to this phenomenon, the quantum states that were pure in nature are converted into mixed states \cite{lidar2013quantum}.

Decoherence is measured with the help of \textit{decoherence times},  i.e. time for which the qubits can be entangled without any loss of information. So, the computation must be completed before the qubits loose information. It depends upon the technology used for qubits. Qubits realized with superconducting circuits, exhibits 100 micro-seconds of decoherence time \cite{van2016path}, and a much larger decoherence time has been reported with trapped ions \cite{fisk1997trapped}. Reliability of a teleportation system is measured with quantum fidelity which is considered as the fundamental figure of merit, and larger the imperfections introduced by decoherence, the lower is the fidelity. Therefore, classical or QEC techniques are extensively adopted to preserve the quantum information against decoherence and imperfections \cite{chandra2017quantum}. Usually, the problem of decoherence is overcome by entanglement distillation \cite{duan2001long}, \cite{chandra2017quantum}, \cite{kurpiers2018deterministic},  which requires an additional level of qubit processing.

\subsection{Imperfect Quantum Teleportation}
\label{sec: Sec. VI-B}

Imperfections are present in all communication modes, and quantum communication is not an exception. These imperfections result from decoherence, as discussed in \textbf{Sec. \ref{sec: Sec. VI-A}}. Quantum communication suffers from environmental interactions and loss of photons carrying qubits. Decoherence also affects the quantum teleportation process as well as entanglement generation and distribution process. However, other operations that are applied to quantum states of quantum teleportation process, further introduce imperfections in the teleported qubit \cite{cacciapuoti2019quantum}. These quantum imperfections are multiplicative in nature, therefore, it can be considered that this phenomenon is more reminiscent of the classical fading effect, rather than the classical additive noise caused by the Brownian motion of electrons \cite{williamson1968brownian}.

A lot of research is going on to accurately model the quantum-domain of imperfections, that are capable of taking into consideration the different effects of imperfections on the quantum teleportation process \cite{cacciapuoti2020entanglement}. 

\subsection{Entanglement of Nodes across the Network}
\label{sec: Sec. VI-C}
One of the biggest challenge for \textcolor{black}{QI} is generation and distribution of entanglement between different nodes across the network. It is the central element to the quantum mechanics and is the most interesting phenomena that has no counterpart in classical communication \cite{Kimble2008}. An EPR pair generated by striking laser beam to a crystal (see \textbf{Sec. \ref{sec: Sec. III-A}}), is responsible for entanglement generation and distribution in the quantum teleportation process which is the key functionality of \textcolor{black}{QI}.
The major challenge that is involved in entanglement distribution is, to extend entanglement to nodes that are at a long distance apart as well as storing the states of qubits at the nodes and regularly updating them using quantum codes. This purpose is solved by employing  quantum repeaters \cite{van2013designing}, but despite of all these schemes, extensive research is required in quantum repeaters, quantum memories, QEC codes and entanglement purification and distillation \cite{hu2021long}, so that multiparty entanglement is free of imperfections and decoherence. Although the physics community has conducted in-depth research on the long-distance entanglement distribution in the past 20 years, and the entanglement distribution rate increases with distance, it still poses a key problem in realization of \textcolor{black}{QI}.

\subsection{Design complexity}
\label{sec: Sec. VI-D}
Unlike classical communication where data in packets is sent along with copies of the same, just in case the data is lost or attacked, it is impossible in quantum communication to copy or amplify qubits because of the properties of no-cloning theorem \cite{cacciapuoti2020entanglement} and the postulate of measurement of quantum state. This put some stringent design challenges, that increases the complexity of the design of network functionalities in \textcolor{black}{QI}. If we talk about classical error-correcting codes, they cannot be directly applied to quantum communication, because re-transmission of qubits is not possible. Therefore, QEC codes are required and network functionalities have to be designed accordingly making the system complex. Such design complexity is present in every layer of \textcolor{black}{QI} design starting from functionality layer of error correcting to the layer of medium access control and route discovery, to the layer-4 protocols such as TCP/IP \cite{wehner2018quantum}, \cite{cerf2005protocol}. 

\textcolor{black}{A corollary of no-cloning theorem is no-broadcasting theorem \cite{cacciapuoti2020entanglement}, where quantum information cannot be transmitted to more than one recipients. Therefore, the link layer must be thoughtfully redesigned to take care of design complexity \cite{cacciapuoti2019quantum}.} 

\subsection{ Measurement of Quantum states.}
\label{sec: Sec. VI-E}

As discussed in \textbf{Sec. \ref{sec: Sec. II-A}-1}, a qubit can be in superposition of the two basis states and measuring the quantum state of qubit will collapse it into one of the basis states \cite{furnkranz2020quantum}. 
This means that prior to measurement, what amount of information can be stored and transmitted by a qubit is still unknown. The problem of measurement is best illustrated by "paradox" of Schrödinger's cat \cite{schrodinger1935gegenwartige}. A cat can be alive or dead both at the same time due to superposition. Upon measurement, the answers is always a living cat or a dead cat. How a probabilistic result landed into a definite result is still a debatable issue and lots of hidden information is there behind this paradox. 

\section{FUTURE DIRECTIONS}
\label{sec: Sec. VII}
Here, we present some future perspectives of \textcolor{black}{QI} technology to make this vision a reality.   

\subsection{Evolution of Quantum Processors}
\label{sec: Sec. VII-A}
Quantum computer with great computation capabilities will be required in future \textcolor{black}{QI} technologies, for performing complex computations. They can solve the problems that are practically impossible with state-of-the-art supercomputers. The quantum network is becoming powerful and efficient because of the increase in the number of qubits at the end nodes. To meet these requirements, quantum processors with double-digit qubits are evolving, that will shape the future of quantum networking. This technological development would help boost the economies of the world, as various cloud operators will make fortunes, with BQC protocols maintaining user privacy. Also, quantum computers will be linked to each other through pure entanglement, and communication will take place using quantum teleportation. The \textcolor{black}{QI} might also assist quantum computer networking to reach even greater performance capacities.

\subsection{Implementation of Trusted Repeater Network}
\label{sec: Sec. VII-B}
The repeater network plays an important role in modern networking and therefore, a lot of theoretical and practical research have been established \cite{collins2005quantum}, \cite{dur1999quantum}, \cite{yuan2008experimental}, \cite{briegel1998quantum}, even if they are still at a developing stage \cite{alleaume2014using}, \cite{salvail2010security}. Implementing repeater networks requires number of intermediary nodes that are entangled to each other so that entanglement is generated between two far away end nodes with the phenomena of \textit{entanglement swapping} as discussed in \textbf{Sec. \ref{sec: Sec. IV-B}-2}. Each pair of adjacent repeater nodes exchange the encrypted keys using QKD which allows the far away end nodes to generate their own keys with a assumption that nodes should be trusted. \textcolor{black}{But, until recently in Bristol work \cite{joshi2020trusted}, trust-free nodes are also proved scalable in a city wide quantum network utilizing entanglement-based QKD. Thus, trusted nodes assumption can be avoided in three ways, firstly by use of entanglement-based QKD \cite{liu2020entanglement}, measurement-device independent QKD \cite{lo2012measurement}, \cite{tang2016measurement}, \cite{hu2018measurement}, multi-path communication techniques such as quantum network coding \cite{hayashi2007quantum} and by use of quantum repeaters \cite{briegel1998quantum}.} But still, trusted repeater is a hot topic of research that will be implemented along with trust-free nodes to overcome distance limitations between QKD nodes.

\subsection{Quantum Error Correcting Codes}
\label{sec: Sec. VII-C}

The future \textcolor{black}{QI} technologies will require advanced QEC codes that works on the basis of encoding qubits carrying quantum information in a unique way that mitigates the influence of noise, decoherence and other environmental factors \cite{knill1997theory}. After that, quantum information is decoded to retrieve the original quantum state \cite{babar2018duality}. The Decoherence Free Subspace (DFS) stores data in multiple qubits instead of storing data in a single qubit, which selects specific aspects of the system that are less affected by one or more important environmental factors \cite{lidar1998decoherence}. In this way, DFS stores data in a subspace of the Hilbert space related to quantum systems that are least influenced by the interaction of its system with the surrounding environment \cite{suter2016colloquium}, \cite{koch2016controlling}. But, identifying DFS for complex quantum systems is extremely difficult \cite{koch2016controlling}, which requires more advanced DFS techniques such as Dynamic Decoupling (DD) \cite{viola1999dynamical}, \cite{viola1998dynamical}, where radio-frequency pulses are applied as external interaction to manipulate the non-unitary component of quantum system \cite{viola1998dynamical}. Nevertheless, the robust design of the sequence provided by the DD technique suppresses the imperfections in the experiments upto a greater extent \cite{suter2016colloquium}. This implies that more advanced techniques are required and further research is needed in QEC codes.

\subsection{Exploring Interface between DV and CV}
\label{sec: Sec. VII-D}
Discrete Variable (DV) and Continuous Variable (CV) describes the states where quantum states can be represented in the quantum communications \cite{nielsen2002quantum}, \cite{braunstein2005quantum}. In DV states, data are represented by discrete features such as the polarization of single photons \cite{bennett1993teleporting}, which can be detected by single-photon detectors. Also, the information is represented by finite number of basis states, such as ‘qubit' which is the the standard unit of DV quantum states. Alternative to this approach are the CV quantum states that were introduced in \cite{bennett1992quantum}, \cite{ralph1999continuous}. Here the information is encoded onto the optical field defined by quadrature variables that constitutes infinite dimensional hilbert space and is useful for quantum information carriers such as lasers \cite{grosshans2002reverse}. CV states can be detected by highly efficient homodyne or heterodyne detectors having faster transmission rates than single-photon detectors \cite{croal2016free}. Therefore, an extensive research in theoretical and experimental domains is needed at the interface of CV and DV states, that will extract the best features from both these technologies to exploit the best of both. 

\subsection{Integrating Qubits with other technologies}
\label{sec: Sec. VII-E}

Many powerful computers that are built use different synergies of systems such as bits, neurons, and qubits for computation. Summit supercomputer that has been built by IBM, has a peak performance of 200,000 teraflops, and is the world's most powerful machine that is built on the ‘bits + neurons' platform. This supercomputer can solve complex computation problems and tasks related to AI. Conclusively, supercomputers based upon ‘bits + neurons' scheme expedite technologically suitable workloads and deliver novel scientiﬁc insights. Neurons can also be combined with qubits, and together they can build a quantum computer with neuron-inspired machine learning algorithms to derive a quantum advantage over classical computation \cite{havlivcek2019supervised}. Qubits along with bits can be another combination based on which quantum processors can be developed \cite{peruzzo2014variational}, that has capabilities well beyond the reach of classical computations. Therefore, exploring possible synergies of ‘bits + qubits + neurons' will shape the future of computing. With this, the capabilities of ‘bits + qubits + neurons' must be extensively researched and will remain a topic of active research in coming years. However, some advanced computing systems and use cases are already operating at the intersection between the pairs of these computing methods.

\subsection{Exploration of Solid-state memories}
\label{sec: Sec. VII-F}
Solid-state qubits are formed by electron spins \cite{loss1998quantum} followed by superconducting circuits \cite{kjaergaard2020superconducting}, comprising of Josephson junctions, interconnects, and passive elements, that are designed to behave as a quantum mechanical two-stage system with good isolation. Solid state quantum memories are more advantageous and attractive as they can maintain its coherence even at cryogenic temperatures. For example, crystalline-solid spin ensembles formed by implanting defects in lattice; such as implanting Nitrogen-Vacancy (NV) centers into diamonds, or by rare-earth-doped crystals, can be proved coherent for hours at cryogenic temperatures. Hence, comprehensive efforts are needed to explore the union of superconducting processors and solid-state quantum memories to enhance the performance of transmission and generation of microwave photons. Accordingly, steps must be taken to investigate on-chip teleportation between a superconducting qubit and NV in a local quantum memory. If these integrations are successful, this hybrid technology would become the most encouraging design to be scaled up into a comprehensive quantum network. Therefore, employing such hybrid quantum computers as end nodes in quantum networking, the next 5-10 years could see the implementation of a hybrid-technology \textcolor{black}{QI} globally.

\subsection{Quantum-classical synergy}
\label{sec: Sec. VII-G}
\textcolor{black}{A future QI will soon be developed owing to the launch of various quantum satellites paving the way for long-distance entanglement and QKD with eavesdropping detection. Quantum communication links will require already mature fiber channels for the exchange of classical data required for the quantum teleportation process \cite{chen2021integrated}. QKD also requires classical channels for information reconciliation and privacy amplification, to achieve higher key rates and reduced latencies over noisy communication channels \cite{mehic2020error}. A fully secure QKD based channels will ultimately connect classical devices such as mobiles, tablets, smart wearables, satellites, smart vehicles with utmost security. A secure quantum cloud will exist, that will take care of user privacy and its data through BQC. Thus, future QI will work in synergy with the classical internet giving rise to a universe of applications with excellent security.}

\section{CONCLUSION}

In this survey, we have discussed the quantum internet and its technological advances that cover the basic concepts to understand this novel technology. We have introduced the readers to the basic physics underlying this technology and the unmatched security and privacy that it offers due to laws of inherently secure quantum mechanics. The future quantum internet will exchange security based on various \textcolor{black}{quantum key distribution} algorithms and will enables networking of classical and quantum nodes that are located remotely. The most important and challenging requirement for quantum internet is quantum teleportation based on entanglement generation and distribution between a network of far-away nodes. For this, quantum repeaters are employed that generate entanglement between far-away nodes with the help of entanglement swapping among pairwise intermediary nodes, where the quantum information is stored in local quantum memories. The quantum information is fragile, which means that environmental effects such as decoherence, where quantum information is lost with time, will negatively impact quantum communication. Due to decoherence, quantum states lose their properties and enters the world of classical physics. Therefore, realization of the quantum internet technology will come with lots of design challenges and complexities. Consequently, multidisciplinary efforts will be required that will pave the way for the realization of a full-fledged quantum internet. A breakthrough has been achieved since the launch of the quantum satellites that have achieved QKD between two far-away ground stations, which has motivated various research groups to speed up the efforts to build a full-blown quantum internet in near future.

\section*{Acknowledgment}

This material is based upon work supported by the grant received under Erasmus+ KA107 programme (ICM)  (Key Action 1, Higher education agreement between Politecnico di Milano, Italy and Thapar Institute of Engineering and Technology, Patiala, Agreement No. 2018-1-IT02-KA107-047412).

\vspace{-0.3cm}
\bibliographystyle{IEEEtran}
\bibliography{Bibliography.bib}

\vskip -2\baselineskip plus -1fil

\begin{IEEEbiographynophoto}{Amoldeep Singh} is Research Scholar in Electronics and Communication Department at Thapar Institute of Engineering and Technology, Patiala, India. He has received his ME degree in Electronics (VLSI design) from Punjab Engineering college, Chandigarh. He is an exchange student at Politecnico di milano, Milan, Italy and is associated with Erasmus KA-107 project. His research interests includes Quantum communication, Quantum networking, VLSI Interconnects, Optical Interconnects. 
\end{IEEEbiographynophoto}
\vskip -2\baselineskip plus -1fil
\vspace{-0.25cm}
\begin{IEEEbiographynophoto}{Kapal Dev} is Senior Researcher at Munster Technological University, Ireland. Previously, he was a Postdoctoral Research Fellow with the CONNECT Centre, School of Computer Science and Statistics, Trinity College Dublin (TCD). Previously, he worked as 5G Junior Consultant and Engineer at Altran Italia S.p.A, Milan, Italy on 5G use cases. He worked as Lecturer at Indus university, Karachi back in 2014. He is also working for OCEANS Network as Head of Projects to manage OCEANS project processes and functions to improve efficiency, consistency and best practice integration

He was awarded the PhD degree by Politecnico di Milano, Italy under the prestigious fellowship of Erasmus Mundus funded by European Commission. His education Profile revolves over ICT background i.e. Electronics (B.E and M.E), Telecommunication Engineering (PhD) and Post-doc (Fusion of 5G and Blockchain). His research interests include Blockchain, 6G Networks and Artificial Intelligence. He is very active in leading (as Principle Investigator) Erasmus + International Credit Mobility (ICM) and Capacity Building for Higher Education and H2020 Co-Fund projects. Received a few Million Euros funding and few are in review as well as in the writing phase. 

He is evaluator of MSCA Co-Fund schemes, Elsevier Book proposals and top scientific journals and conferences including IEEE TII, IEEE TITS, IEEE TNSE, IEEE IoT, IEEE JBHI, FGCS, COMNET, TETT, IEEE VTC, WF-IoT. TPC member of IEEE BCA 2020 in conjunction with AICCSA 2020, ICBC 2021, DICG Co-located with Middleware 2020 and FTNCT 2020. He is also serving as Associate Editor in Wireless Networks, IET Quantum Communication, IET Networks, and Review Editor in Frontiers in Communications and Networks. He is also serving as Guest Editor (GE) in COMCOM (I.F: 2.8), GE in COMNET (I.F 3.11), Lead chair in one of CCNC 2021 workshops, and editing a book for CRC press.  He  is  a  member  of  the  ACM, IEEE,  and  actively involved  with  various  working  groups  and committees  of  IEEE  and  ACM  related  to 5G and beyond,  Blockchain  and Artificial Intelligence.
\end{IEEEbiographynophoto}
\vskip -2\baselineskip plus -1fil
\vspace{-0.25cm}
\begin{IEEEbiographynophoto}
{Harun Siljak} received his bachelor and master degrees in Automatic Control and Electronics from the University of Sarajevo in 2010 and 2012, respectively and his PhD degree in Electrical and Electronics Engineering from the International Burch University, Sarajevo, in 2015. He is an assistant professor at the School of Engineering, Trinity College Dublin. During his earlier Marie Curie fellowship at TCD, he merged complex systems science and reversible computation in a new toolbox for wireless communications, mainly for control and optimisation of large antenna arrays. The tools developed during the project found their application in quantum computation and communications, as well as molecular and neuronal communications. He serves as an associate editor for the EURASIP Journal on Wireless Communications and Networking, and science communication officer for the Western Balkans Chapter of Marie Curie Alumni Association and MAT-DYN-NET COST Action.
\end{IEEEbiographynophoto}
\vskip -2\baselineskip plus -1fil
\vspace{-0.25cm}
\begin{IEEEbiographynophoto}
{Hem Dutt Joshi} received the B.Tech degree in ECE from Barkatullah University, Bhopal, India in 1999, the ME degree in CCN from MITS, Gwalior in 2004 and the PhD degree from JUET, Guna, India in 2012. He worked as Assistant Professor in JUET, Guna from 2006 to 2013. Currently, he is working as Associate Professor in the department of ECE, TIET, Patiala. Dr Joshi also worked as visiting research fellow in the year 2019 at CONNECT, the Science Foundation Ireland Research Centre for Future Networks, at Trinity College Dublin, Ireland. Dr. Joshi is involved very extensively in research regarding high data rate wireless communication systems and signal processing. He has more than 15 years of teaching and research experience. His research interests include wireless communication, OFDM system, MIMO–OFDM, signal processing for wireless communication and 5G communication system.

\end{IEEEbiographynophoto}
\vskip -2\baselineskip plus -1fil
\vspace{-0.25cm}
\begin{IEEEbiographynophoto}{Maurizio Magarini} received the M.Sc. and Ph.D. degrees in electronic engineering from the Politecnico di Milano, Milan, Italy, in 1994 and 1999, respectively. In 1994, he was granted the TELECOM Italia (now TIM) scholarship award for his M.Sc. Thesis. He worked as a Research Associate in the Dipartimento di Elettronica, Informazione e Bioingegneria at the Politecnico di Milano from 1999 to 2001. From 2001 to 2018, he was an Assistant Professor in Politecnico di Milano where, since June 2018, he has been an Associate Professor. From August 2008 to January 2009 he spent a sabbatical leave at Bell Labs, Alcatel-Lucent, Holmdel, NJ. His research interests are in the broad area of communication and information theory. Topics include synchronization, channel estimation, equalization and coding applied to wireless and optical communication systems. His most recent research activities have focused on molecular communications, massive MIMO, study of waveforms for 5G cellular systems, wireless sensor networks for mission critical applications, and wireless networks using UAVs and high-altitude platforms. He has authored and coauthored more than 100 journal and conference papers. He was the co-recipient of four best paper awards. He is Associate Editor of IEEE Access and IET Electronics Letters and a member of the Editorial Board of Nano Communication Networks (Elsevier) and MDPI Telecom. He has been involved in several European and National research projects. 
\end{IEEEbiographynophoto}
\end{document}